\documentclass[review,12pt]{elsarticle}
\usepackage[margin=1in]{geometry}

\usepackage{type1cm}
\usepackage{makeidx}
\usepackage{graphicx}
\usepackage{multicol}
\usepackage[bottom]{footmisc}
\usepackage{amsmath,amssymb,amscd}
\usepackage{enumitem}
\usepackage[
colorlinks=false,
citecolor=black,
urlcolor=black,
linktocpage,
pagebackref
]{hyperref}
\usepackage{comment}
\usepackage{pgfplots}
\usepackage{algorithmic}
\usepackage[ruled,vlined,linesnumbered]{algorithm2e}
\usepackage{soul}

\usepackage{listings}
\definecolor{darkgreen}{rgb}{0.0, 0.5, 0.0}
\definecolor{mGreen}{rgb}{0,0.6,0}
\definecolor{mGray}{rgb}{0.5,0.5,0.5}
\definecolor{mPurple}{rgb}{0.58,0,0.82}
\definecolor{backgroundColour}{rgb}{0.95,0.95,0.92}
\definecolor{code}{rgb}{0.7, 0, 0.4}

\newcommand{\cred}[1]{{\color{black}{#1}}}
\newcommand{\cblue}[1]{{\color{black}{#1}}}
\newcommand{\cgreen}[1]{{\color{black}{#1}}}

\SetKwRepeat{Do}{do}{while}

\def\bx{\mathbf{x}}   
\def\bdx{{\Delta \bx}}
\def\lim{{\xi}}    

\def\ra{$r-$adaptivity}

\def\bw{\bar{w}}

\def\ws{$w_{\sigma}$}

\journal{Journal}

\begin{document}

\begin{frontmatter}

\title{High-Order Mesh Morphing for Boundary and Interface Fitting to Implicit Geometries}

\author{Jorge-Luis Barrera \fnref{llnl}}
\author{Tzanio Kolev \fnref{llnl}}
\author{Ketan Mittal \fnref{llnl,corr}}
\author{Vladimir Tomov \fnref{llnl}}
\fntext[llnl]
{Lawrence Livermore National Laboratory, 7000 East Avenue, Livermore, CA 94550}
\fntext[corr]
{Corresponding author, mittal3@llnl.gov}
\tnotetext[l_title]
{Performed under the auspices of the U.S. Department of Energy under
Contract DE-AC52-07NA27344 and was supported by the LLNL-LDRD Program under Project No 22-ERD-023 (LLNL-JRNL-837413).}

\address{}
\begin{abstract}
\cred{
We propose a method that morphs high-orger meshes such that their boundaries and interfaces coincide/align with implicitly defined geometries.}
Our focus is particularly on the case when the target surface is prescribed as the zero isocontour of a smooth discrete function. Common examples of this scenario include using level set functions to represent material interfaces in multimaterial configurations, and evolving geometries in shape and topology optimization. The proposed method formulates the mesh optimization problem as a variational minimization of the sum of a chosen mesh-quality metric using the Target-Matrix Optimization Paradigm (TMOP) and a penalty term that weakly forces the selected faces of the mesh to align with the target surface.
The distinct features of the method are use of a source mesh to represent the level set function with sufficient accuracy, and adaptive strategies for setting the penalization weight and selecting the faces of the mesh to be fit to the target isocontour of the level set field. We demonstrate that the
proposed method is robust for generating boundary- and interface-fitted meshes for curvilinear domains using different element types in 2D and 3D.
\end{abstract}

\begin{keyword}
High-order \sep  Implicit meshing \sep Mesh morphing \sep $r$-adaptivity \sep Finite elements \sep TMOP
\end{keyword}
\end{frontmatter}

\section{Introduction}
\label{sec_intro}
High-order finite element (FE) methods are becoming increasingly relevant in
computational science and engineering disciplines due to their potential for better simulation accuracy and
favorable scaling on modern architectures \cite{Fischer2002,fischer2020scalability,MFEM2020, CEED2021}.
A vital component of these methods is high-order computational meshes for discretizing the geometry.
Such meshes are essential for achieving optimal convergence rates on domains
with curved boundaries/interfaces, symmetry preservation, and alignment with
the key features of the flow in moving mesh simulations
\cite{Shephard2011, Dobrev2012, Boscheri2016}.

\cred{
To fully benefit from high-order geometry representation,
however, one must be able to control the quality and
adapt the properties of a high-order mesh.
Two common requirements for mesh optimization methods are
(i) to fit certain mesh faces to a given surface representation,
and (ii) to perform tangential node movement along a mesh surface.
This paper is concerned with these two requirements, in the particular case
when the surface representation is a discrete (or implicit) function.
Common examples of this scenario include use of level set functions to represent curvilinear domains as a combination of geometric primitives in
Constructive Solid Geometry (CSG) \cite{requicha1977constructive}, material
interfaces in multimaterial configurations \cite{Osher1994}, and evolving
geometries in shape and topology optimization
\cite{sokolowski1992introduction,allaire2014shape,hojjat2014vertex},
amongst other applications.}

Boundary conforming high-order meshes are typically generated by starting with a conforming linear mesh that is projected to a higher order space before the mesh faces are curved to fit the boundary \cite{xie2013generation,fortunato2016high,moxey2015isoparametric,gargallo2016distortion,toulorge2016optimizing,poya2016unified,ruiz2022automatic}.
\cred{In context of boundary fitting, a closely related work is Rangarajan's method for tetrahedral meshes \cite{rangarajan2019algorithm} where an immersed linear mesh is trimmed, projected to a surface defined using a point-cloud, and smoothed to generate a high-order boundary fitted mesh. Other related approaches are Mittal's distance function-based approach for tangential relaxation during optimization of initially fitted high-order meshes \cite{mittal2019mesh} and the \textit{DistMesh} algorithm where Delaunay triangulations are aligned to implicitly defined domains using force balance.
Note that we are interested in generating boundary fitted meshes using mesh morphing because it offers a way to use existing finite element framework for adapting meshes to the problem of interest. An alternative is classic mesh generation, which is usually a pre-processing step, and has been an area of interest to the meshing community for decades; summarizing different mesh generation techniques is beyond the scope of this work and the reader is referred to \cite{bommes2013quad,baker2005mesh,lo2014finite} for a review on this topic.}

For generating interface fitted meshes, existing methods mainly rely on topological operations where the input mesh is split across the interface to generate an interface conforming mesh \cite{chen2017interface,mmgplatform}.
Some exceptions are Ohtake's method for adapting linear triangular surface meshes to align with domains with sharp features \cite{ohtake2002dual}
and Le Goff's method for aligning meshes to interfaces prescribed implicitly using volume fractions \cite{le2017volume}.
Barring \cite{persson2004simple,rangarajan2019algorithm,abdelkader2020vorocrust,ohtake2002dual,le2017volume},
existing methods mainly rely on an initial conforming meshing for boundary fitting and
on topological operations such as splitting for interface fitting.

We propose a boundary and interface fitting method for high-order meshes that is  algebraic and seldom requires topological operations,
extends to different element types (quadrilaterals/triangles/hexahedra/tetrahedra)
in 2D and 3D, and works for implicit parameterization of the target surface using discrete finite element (FE) functions.
We formulate the implicit meshing challenge as a mesh optimization
problem where the objective function is the sum of a
chosen mesh-quality metric using the Target-Matrix Optimization Paradigm (TMOP) \cite{TMOP2019SISC, Knupp2012} and a penalty
term that weakly forces nodes of selected faces of a mesh to align with
the target surface prescribed as the zero level set of a discrete
function. Additionally, we use an adaptive strategy to choose element faces/edges for alignment/fitting and set the penalization weight, to ensure robustness of the method for nontrivial curvilinear boundaries/interfaces.
We also introduce the notion of a source mesh that can be used to accurately represent the level set with sufficient accuracy.
This mesh is decoupled from the mesh being optimized, which allows to
represent the domain of interest with higher level of detail.
This approach is crucial for cases where the target boundary is
beyond the domain of the mesh being optimized or the input
mesh does not have sufficient resolution around the zero level set.

The remainder of the paper is organized as follows.
In Section \ref{sec_prelim} we review the basic TMOP components and
our framework to represent and optimize high-order meshes.
The technical details of the proposed method for surface fitting and tangential
relaxation are described in Section \ref{sec_bif}.
Section \ref{sec_results} presents several academic tests that
demonstrate the main features of the methods,
followed by conclusions and direction for future work in Section \ref{sec_concl}.

\section{Preliminaries}
\label{sec_prelim}
In this section, we describe the key notation and our prior work
that is relevant for understanding our newly developed boundary-
and interface-fitting method.
\subsection{Discrete Mesh Representation}
\label{sec_mesh}

In our finite element based framework, the domain $\Omega \in \mathbb{R}^d$, $d=\{2,3\}$, is
discretized as a union of $N_E$ curved mesh elements, each of order $p$.  To obtain a
discrete representation of these elements, we select a set of scalar basis
functions $\{ \bar{w}_i \}_{i=1}^{N_p}$, on the reference element $\bar{E}$.
For example, for tensor-based elements (quadrilaterals in 2D, hexahedra in 3D),
we have $N_p = (p+1)^d$, and the basis spans the space
of all polynomials of degree at most $p$ in each variable, denoted by $Q_p$.
These $p$th-order basis functions are typically chosen to be Lagrange
interpolation polynomials at the Gauss-Lobatto nodes of the reference element.
The position of an
element $E$ in the mesh $\mathcal{M}$ is fully described by a matrix
$\mathbf{x}_E$ of size $d \times N_p$ whose columns represent the coordinates
of the element {\em control points} (also known as {\em nodes} or element {\em degrees of freedom}).
Given $\mathbf{x}_E$ and the positions $\bar{x}$ of the
reference element $\bar{E}$, we introduce the map
$\Phi_E:\bar{E} \to \mathbb{R}^d$ whose image is the geometry of the physical element $E$:
\cgreen{
\begin{equation}
\label{eq_x}
x(\bar{x}) =
   \Phi_E(\bar{x}) \equiv
   \sum_{i=1}^{N_p} \mathbf{x}_{E,i} \bw_i(\bar{x}),
   \qquad \bar{x} \in \bar{E}, ~~ x=x(\bar{x}) \in E,
\end{equation}
}
where $\mathbf{x}_{E,i}$ denotes the $i$-th column of $\bx_E$, i.e.,
the $i$-th node of element $E$.
Throughout the manuscript, $x$ will denote the position function defined
by \eqref{eq_x}, while bold $\bx$ will denote the global vector of all
node coordinates.

\subsection{Geometric Optimization and Simulation-Based \ra\ with TMOP}
\label{sec_ra}

The input of TMOP is the user-specified transformation matrix $W$, from
reference-space $\bar{E}$ to target element $E_t$, which represents the
ideal geometric properties desired for every mesh point. Note that after
discretization, there will be multiple input transformation matrices $W$ -- one
for every quadrature point in every mesh element.
The construction of this transformation is guided by the fact that any Jacobian
matrix can be written as a composition of four \cblue{geometric} components:
\begin{equation}
\label{eq_W}
W = \underbrace{\zeta}_{\text{[volume]}} \underbrace{R}_{\text{[rotation]}}
\underbrace{Q}_{\text{[skewness]}} \underbrace{D}_{\text{[aspect ratio]}}.
\end{equation}
A detailed discussion on the construction of matrices associated with these geometric components and on how TMOP's target construction methods encode geometric information into the target
matrix $W$ is given by Knupp in \cite{knupp2019target}. Various
examples of target construction for different mesh adaptivity goals are given in
\cite{TMOP2019SISC,TMOP2020CAF,TMOP2021EWC}.

\begin{figure}[tb!]
\centerline{
  \includegraphics[width=0.5\textwidth]{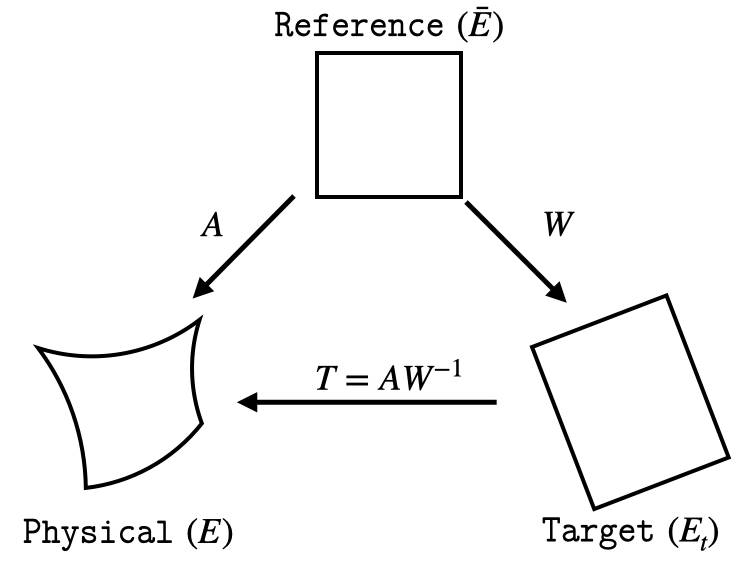}}
\caption{Schematic representation of the major TMOP matrices.}
\label{fig_tmop}
\end{figure}

Using \eqref{eq_x}, the Jacobian of the mapping $\Phi_E$ at any reference point
\cblue{$\bar{x} \in \bar{E}$} from the reference-space coordinates to
the current physical-space coordinates is
\begin{equation}
\label{eq_A}
  A(\bar{x}) = \frac{\partial \Phi_E}{\partial \bar{x}} =
    \sum_{i=1}^{N_w} \mathbf{\bx}_{E,i} [ \nabla \bar{w}_i(\bar{x}) ]^T \,.
\end{equation}
\cgreen{In this manuscript, we assume that all the elements in the initial mesh are not inverted, i.e. $\text{det}(A) > 0\,\, \forall \bar{x} \in \bar{E}$.}
Combining \eqref{eq_A} and \eqref{eq_W}, the transformation from the
target coordinates to the current physical coordinates (see Fig. \ref{fig_tmop})
is
\begin{equation}
\label{eq_T}
T = AW^{-1}.
\end{equation}

With the target transformation $W$ defined in the domain, we next specify a
\cred{mesh quality metric $\mu(T)$ that compares the
transformations $A$ and $W$ in terms of the geometric parameters of interest}. \cblue{For example, $\mu_2=\frac{\mid T \mid^2}{2\tau}-1$
is a \emph{shape} metric\footnote{We follow the metric numbering in \cite{Knupp2020,knupp2022geometric}.} that depends on the skewness and aspect ratio components, but
is invariant to orientation/rotation and volume. \cblue{Here, $|T|$ and $\tau$ are the Frobenius norm and determinant of $T$, respectively}. Similarly,
$\mu_{77}=\frac{1}{2}(\tau-\frac{1}{\tau})^2$ is a
\emph{size} metric that depends only on the volume of the element.
We also have \emph{shape}$+$\emph{size} metrics such as $\mu_{80} =\gamma \mu_2 + (1-\gamma) \mu_{77}$
that depend on volume, skewness and aspect ratio,
but are invariant to orientation/rotation.}
Note that the mesh quality
metrics are defined such that they evaluate to 0 for an identity
transformation, i.e. $\mu(T) = 0$ when $T=I$ ($A=W$).
This allows us to pose the mesh optimization problem as minimization of $\mu(T)$,
amongst other advantages \cite{Knupp2020}.

The quality metric $\mu(T)$ is used to define the TMOP objective
function for \ra
\begin{equation}
\label{eq_F_full}
  F(\bx) = \sum_{E \in \mathcal{M}} F_E(\bx_E)
         = \sum_{E \in \mathcal{M}} \int_{E_t} \mu(T(x)) dx_t,
\end{equation}
where $F$ is a sum of the TMOP objective function for each element in the mesh ($F_E$),
and $E_t$ is the target element corresponding to the element $E$.
In \eqref{eq_F_full}, the integral is computed as
\begin{equation}
\label{eq_vm}
  \sum_{E \in \mathcal{M}} \int_{E_t} \mu(T(x_t)) dx_t =
  \sum_{E \in \mathcal{M}} \sum_{x_q \in E_t}
                           w_q\,\det(W(\bar{x}_q))\, \mu(T(x_q)),
\end{equation}
where $\mathcal{M}$ is the current mesh with $N_E$ elements, $w_q$ is the
quadrature weight, and the position $x_q$ is the image of the reference
quadrature point location $\bar{x}_q$ in the target element $E_t$.

Optimal node locations are determined by minimizing the TMOP objective function
\eqref{eq_F_full}.
This is performed by solving $\partial F(\bx) / \partial \bx = 0$
using the Newton's method where we
improve the mesh degrees-of-freedom (nodal positions) as
\begin{eqnarray}
\label{eq_r_adaptivity_solve}
\bx_{k+1} = \bx_k - \alpha \mathcal{H}^{-1}(\bx_k)\mathcal{J}(\bx_k).
\end{eqnarray}
Here, $\bx_k$ refers to the nodal positions at the $k$-th Newton iteration during
\ra, $\alpha$ is a scalar determined by a line-search procedure, and
$\mathcal{H}(\bx_k)$ and $\mathcal{J}(\bx_k)$ are the
Hessian \cblue{($\partial^2 F(\bx_k) / \partial \bx_j\partial \bx_i$)}
and the gradient ($\partial F(\bx_k) / \partial \bx_i$),
respectively, associated with the TMOP objective function. The line-search
procedure requires that $\alpha$ is chosen such that $F(\bx_{k+1}) < 1.2 F(\bx_k)$,
$|\mathcal{J}(\bx_{k+1})| < 1.2 |\mathcal{J}(\bx_k)|$, and
that the determinant of the Jacobian of the transformation at each quadrature point
in the mesh is positive, $\det(A(\bx_{k+1})) > 0$.
\cgreen{These line-search constraints have been tuned using many numerical experiments
and have demonstrated to be effective in improving mesh quality.}
For Newton's method, we solve the problem \cblue{$\mathcal{H}(\bx_k) \bdx = \mathcal{J}(\bx_k)$}
using a Krylov subspace method such as the Minimum Residual (MINRES) method
with Jacobi preconditioning;
more sophisticated preconditioning techniques can be found in \cite{Roca2022}. Additionally,
the optimization solver iterations \eqref{eq_r_adaptivity_solve} are
done until the relative $l_2$ norm of the gradient of the objective function
with respect to the current and original mesh nodal positions is below a certain
tolerance $\varepsilon$, i.e., $|\mathcal{J}(\bx)|/|\mathcal{J}(\bx_0)| \leq \varepsilon$.
We set $\varepsilon = 10^{-10}$ for the results presented in the current work.

\begin{figure}[t!]
\begin{center}
$\begin{array}{cc}
\includegraphics[height=0.3\textwidth]{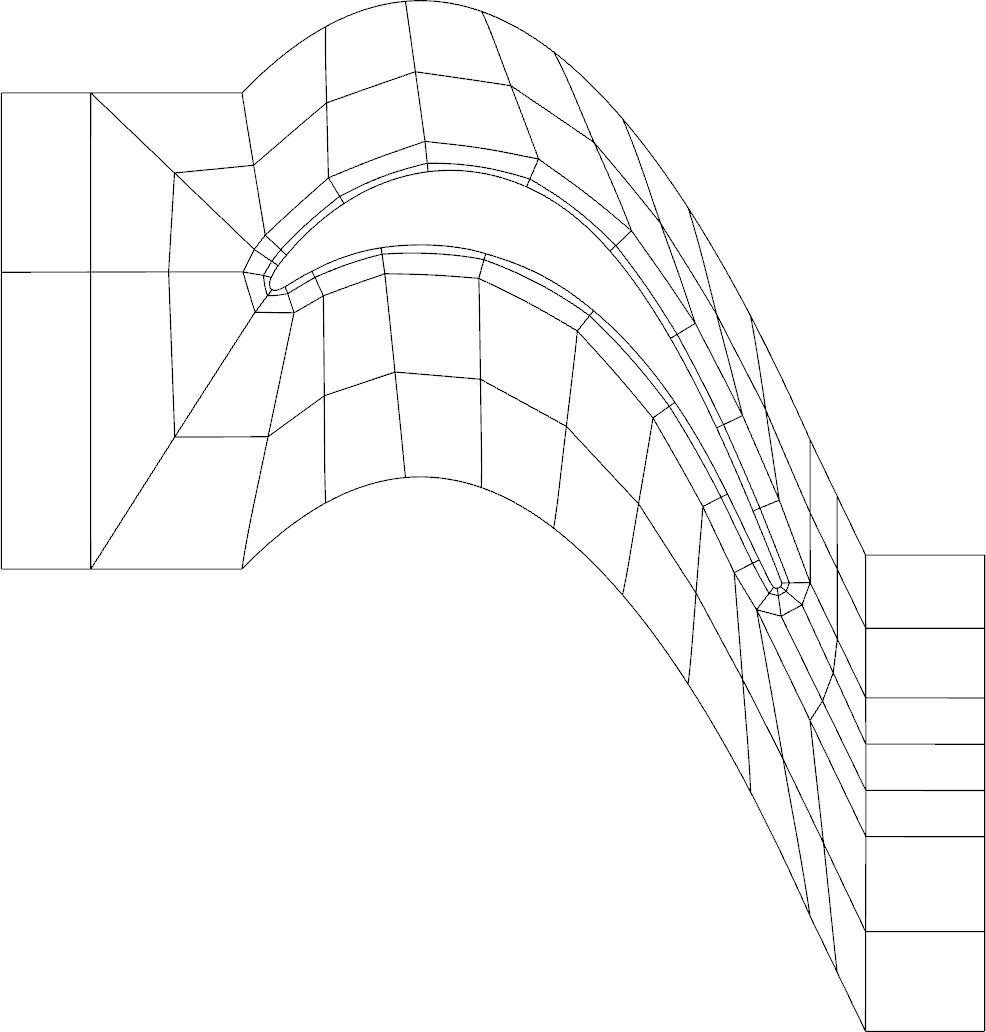} &
\includegraphics[height=0.3\textwidth]{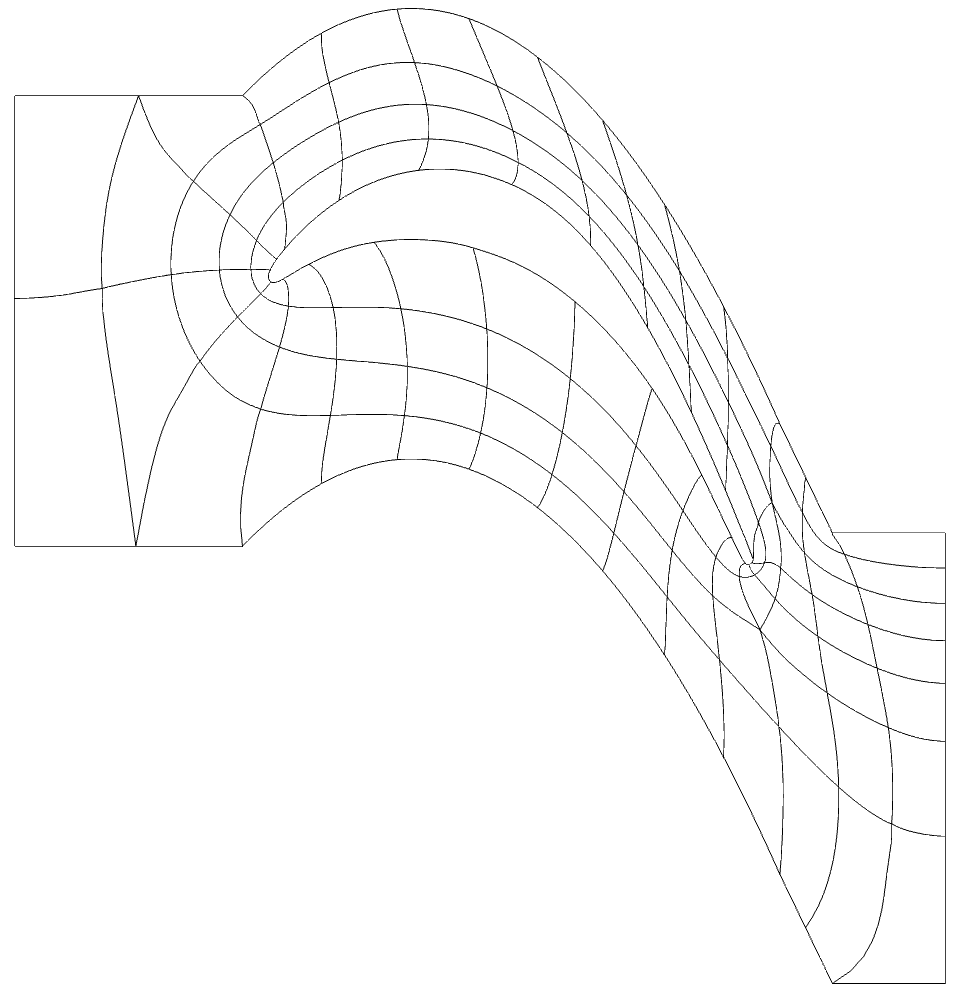}
\vspace{-2mm} \\
\textrm{(a)} & \textrm{(b)} \\
\end{array}$
\end{center}
\vspace{-7mm}
\caption{(a) Original and (b) optimized 4th order meshes for a turbine blade.}
\label{fig_blade}
\end{figure}

Using the approach described in this section, we have demonstrated \ra\ with TMOP
for geometry and simulation-driven optimization; see Fig. \ref{fig_blade}
for example of high-order mesh optimization for a turbine blade using $W=I$ with a shape metric. \cblue{The resulting optimized mesh has elements closer to unity aspect ratio and skewness closer to $\pi/2$ radians in comparison to the original mesh, as prescribed by the target $W=I$.}

\section{Boundary \& Interface Fitting}
\label{sec_bif}

\begin{figure}[tb!]
\begin{center}
$\begin{array}{ccc}
\includegraphics[height=0.3\textwidth]{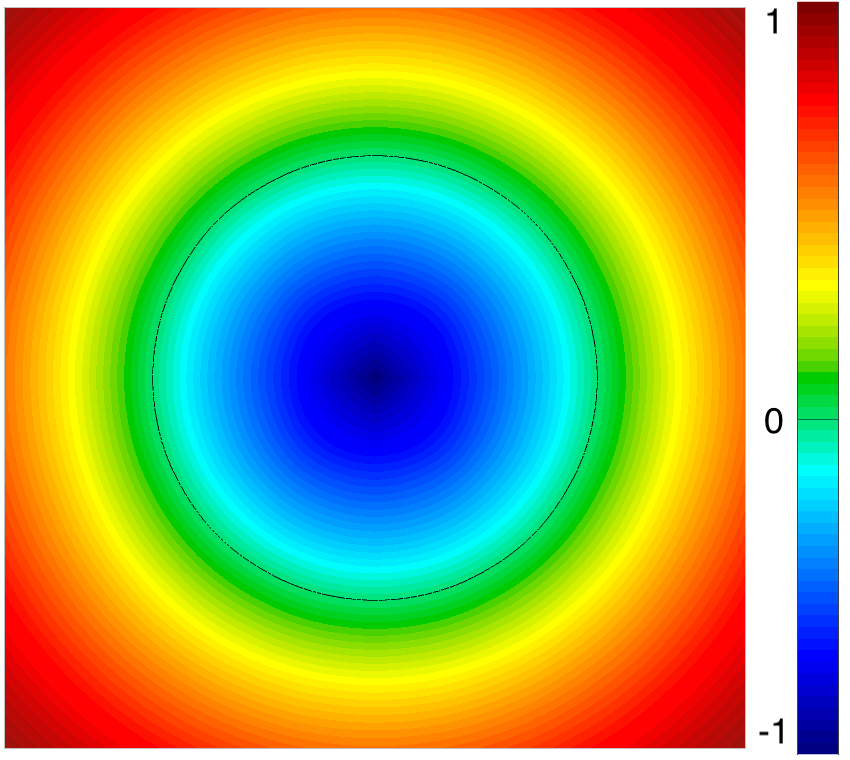} &
\includegraphics[height=0.3\textwidth]{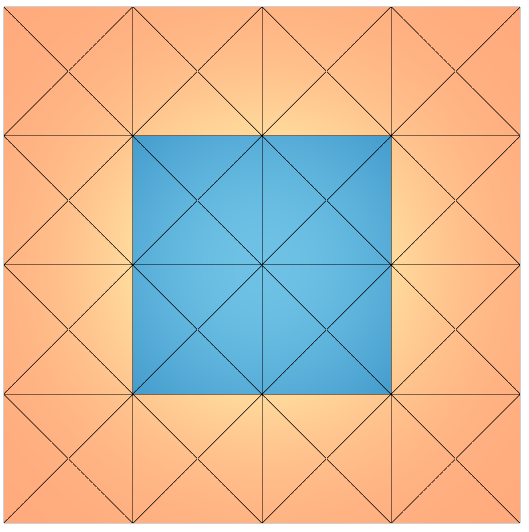} &
\includegraphics[height=0.3\textwidth]{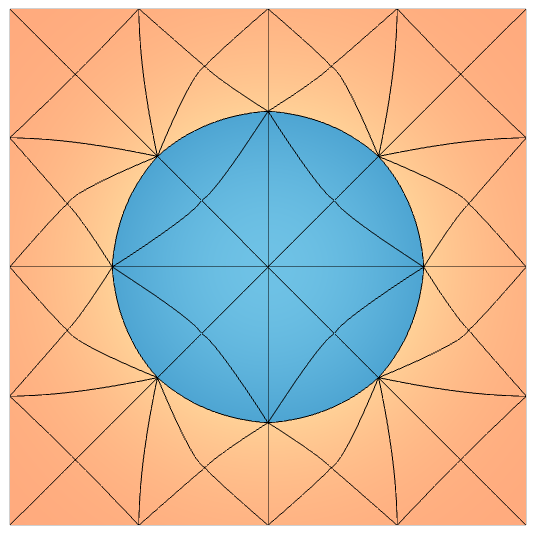}
\vspace{-2mm}\\
\textrm{(a)} &
\textrm{(b)} &
\textrm{(c)} \\
\end{array}$
\end{center}
\vspace{-7mm}
\caption{(a) Level set function $\sigma(x)$, (b) a Cartesian mesh with material interface nodes to be aligned to the zero level set of $\sigma(x)$, and (c) the optimized mesh.}
\label{fig_square_circle}
\end{figure}

Our goal for boundary and interface fitting is to enable alignment of a selected set of mesh nodes to some target surface of interest prescribed as the zero isocontour of a smooth \cgreen{signed-}discrete level set function, $\sigma(x)$.
Figure \ref{fig_square_circle}(a) and (b) show a simple example of a circular interface represented using a level set function and a triangular mesh with multimaterial interface that is to be aligned to the circular interface.
To effect alignment with the zero isocontour of $\sigma(x)$, we modify
the TMOP objective function \eqref{eq_F_full} as:
\begin{equation}
\label{eq_F_full_sigma}
  F(\bx) = \underbrace{\sum_{E \in \mathcal{M}} \int_{E_t}
  \mu(T(x)) dx_t}_{F_{\mu}} +
  \underbrace{w_{\sigma} \sum_{s \in S} \sigma^2(x_s)}_{F_{\sigma}}.
\end{equation}
Here, $F_{\sigma}$ is a penalty-type term that depends on the penalization
weight \ws, the set of nodes $\mathcal{S}$ to be aligned to the level set (e.g., the mesh
nodes discretizing the material interface in Fig. \ref{fig_square_circle}(b)),
and the level set function $\sigma(x)$, evaluated at the positions $x_s$
of all nodes $s \in \mathcal{S}$.
The $F_{\sigma}$ term penalizes the nonzero values of $\sigma(x_s)$ for all
$s \in \mathcal{S}$.
Minimizing this term represents weak enforcement of $\sigma(x_s) = 0$,
only for the nodes in $\mathcal{S}$, while ignoring the values
of $\sigma$ for the nodes outside $\mathcal{S}$.
Minimizing the full nonlinear objective function, $F = F_{\mu} + F_{\sigma}$,
produces a balance between mesh quality and surface fitting.
Note that all nodes are treated together, i.e., the nonlinear solver
makes no explicit separation between volume nodes of $\mathcal{M}$ and the nodes $s \in \mathcal{S} \subseteq \mathcal{M}$ set for fitting.
Furthermore, as there is no pre-determined unique target position for
each node of $\mathcal{S}$, the method naturally allows tangential
relaxation along the interface of interest, so that mesh quality
can be improved while maintaining a good fit to the surface.
Figure \ref{fig_square_circle}(c) shows an example of a triangular mesh
fit to a circular interface using \eqref{eq_F_full_sigma}.
In this example, we use the shape metric $\mu_2$ with equilateral targets
and a constant penalization weight, $w_{\sigma}=10^3$.

The first step in our method is to use a suitable strategy for representing the
level set function with sufficient accuracy (Section \ref{sec_level_set_rep}).
\cgreen{
Next, we determine the set of nodes $\mathcal{S}$ that will be aligned to the zero level set of $\sigma$. For boundary fitting, $\mathcal{S}$ depends on the elements located on the boundary of interest.
For interface fitting, $\mathcal{S}$ is the set of nodes shared between elements with different fictitious material indicators, and we describe our approach for setting the material indicators of elements in Section \ref{sec_marking}.}
Finally, we set the penalization weight \ws\
such that an adequate fit is achieved to the target surface while optimizing the mesh with respect to the quality metric $\mu$ (Section \ref{sec_adaptive_sigma}). The adaptive strategy for setting \ws\ requires us to modify the line-search and convergence criterion of the Newton's method (Section \ref{sec_converge}).
For completeness, the derivatives of $F_{\sigma}$ are discussed in Section \ref{sec_derivatives}. Using various examples, we demonstrate in Section \ref{sec_results} that our method extends to both simplices and hexahedrals/quadrilaterals of any order, and is robust in adapting a mesh interface and/or boundary to nontrivial curvilinear shapes.

\subsection{Level Set Representation}
\label{sec_level_set_rep}

\begin{figure}[b!]
\begin{center}
$\begin{array}{cc}
\includegraphics[height=0.25\textwidth]{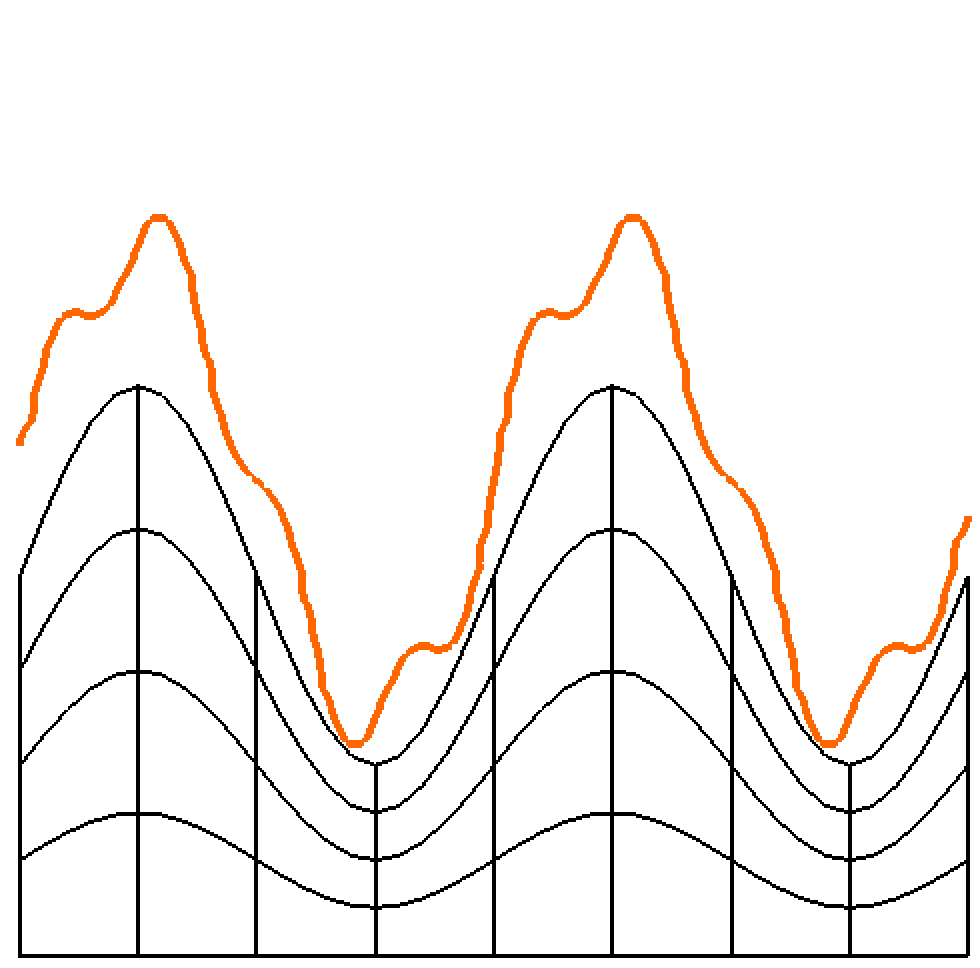} &
\includegraphics[height=0.25\textwidth]{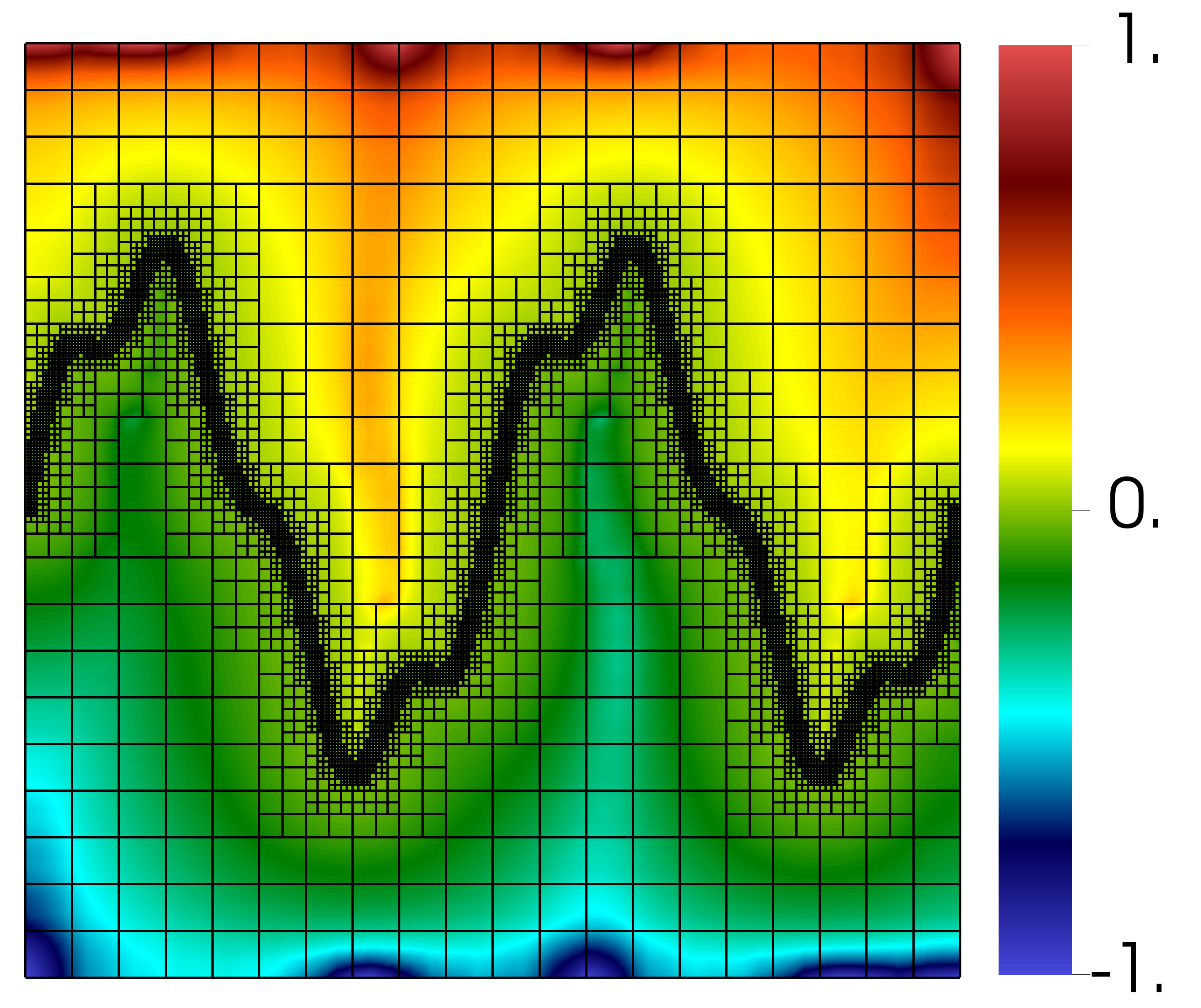} \vspace{-2mm}\\
\textrm{(a)} &
\textrm{(b)} \\
\end{array}$
\end{center}
\vspace{-7mm}
\caption{Using a background mesh to fit discretely prescribed
         domain boundaries. (a) Initial unfitted mesh and target boundary
         curve (orange). (b) The boundary curve is prescribed implicitly as the zero level set of a discrete function on an adaptively refined nonconforming background mesh.}
\label{fig_boundary}
\end{figure}

The following discussion is related to the case when $\sigma(x)$ is a discrete
function so that its values and derivatives can't be computed analytically.
Then the first step in our implicit meshing framework is to ensure that
the level set function $\sigma(x)$ is defined with sufficient accuracy.
A drawback of discretizing $\sigma(x)$ on the mesh being optimized,
$\mathcal{M}$, is that the resulting mesh will be sub-optimal in terms of
mesh quality and interface/boundary fit if
(a) the mesh does not have enough resolution to represent $\sigma(x)$ with sufficient accuracy, especially near the zero level set, or
(b) the target level set is outside the initial domain of the mesh.
For the latter, it's impossible to compute the necessary values and
derivatives of the level set function accurately at the boundary nodes that
we wish to fit, once the mesh moves outside the initial domain;
see \autoref{fig_boundary}(a) for an example.

To address these issues, we introduce the notion of a background/source
mesh ($\mathcal{M}_B$) for discretizing $\sigma(x_B)$, where $x_B$ represents
the positions of the source mesh nodes, as in \eqref{eq_x}.
Since $\mathcal{M}_B$ is independent of $\mathcal{M}$,
we can choose $\mathcal{M}_B$ based on the desired accuracy; \cblue{the maximum error in representing the implicit geometry discretely is bounded by the element size of the background mesh at the location of the zero level set.} Thus, we use adaptive nonconforming mesh refinements \cite{cerveny2019nonconforming} around the zero level set of $\sigma(x_B)$, as shown in Fig. \ref{fig_boundary}(b).
Using a source mesh for $\sigma(x_B)$, however, requires transfer of the
level set function and its derivatives from $\mathcal{M}_B$ to the
nodes $\mathcal{S} \in \mathcal{M}$ prior to each Newton iteration.
This transfer between the source mesh and
the current mesh is done using \emph{gslib}, a high-order interpolation
library \cite{gslibrepo}:
\begin{equation}
\label{eq_interpolation}
\sigma(x) = \mathcal{I}(x, x_B, \sigma(x_B)), ~~
\partial \sigma(x) = \mathcal{I}(x, x_B, \partial \sigma(x_B)), ~~
\partial^2 \sigma(x) = \mathcal{I}(x, x_B, \partial^2 \sigma(x_B)),
\end{equation}
where $\mathcal{I}$ represent the interpolation operator that depends on the
current mesh nodes ($x$), the source mesh nodes ($x_B$), and the source
function $\sigma(x_B)$ or its gradients.
A detailed description of how high-order functions
can be transferred from a mesh to an arbitrary set of points in physical space
using \emph{gslib} is described in Section 2.3 of \cite{mittal2019nonconforming}.

\begin{figure}[b!]
\begin{center}
$\begin{array}{ccc}
\includegraphics[height=0.29\textwidth]{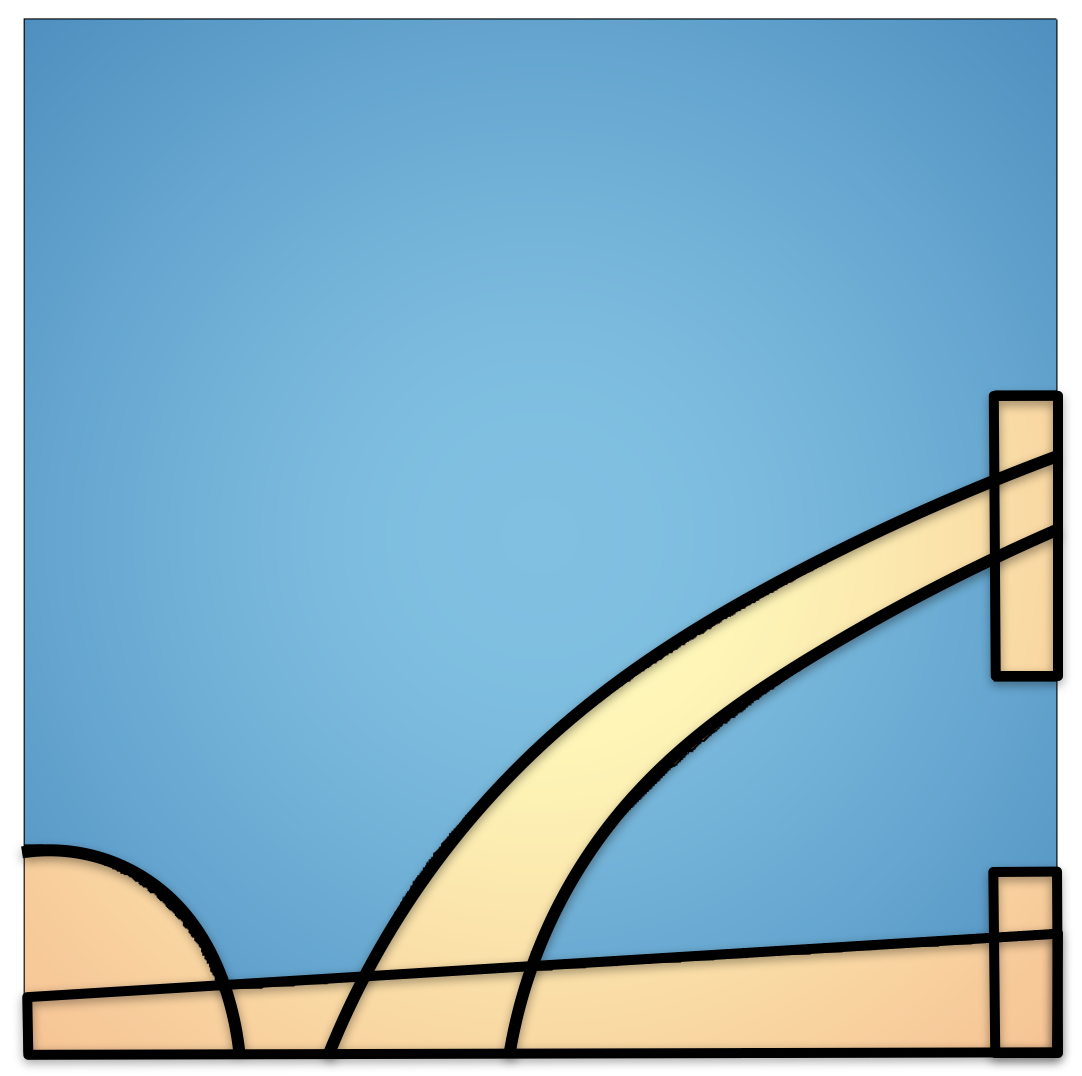} \hspace{2mm} &
   \includegraphics[height=0.29\textwidth]{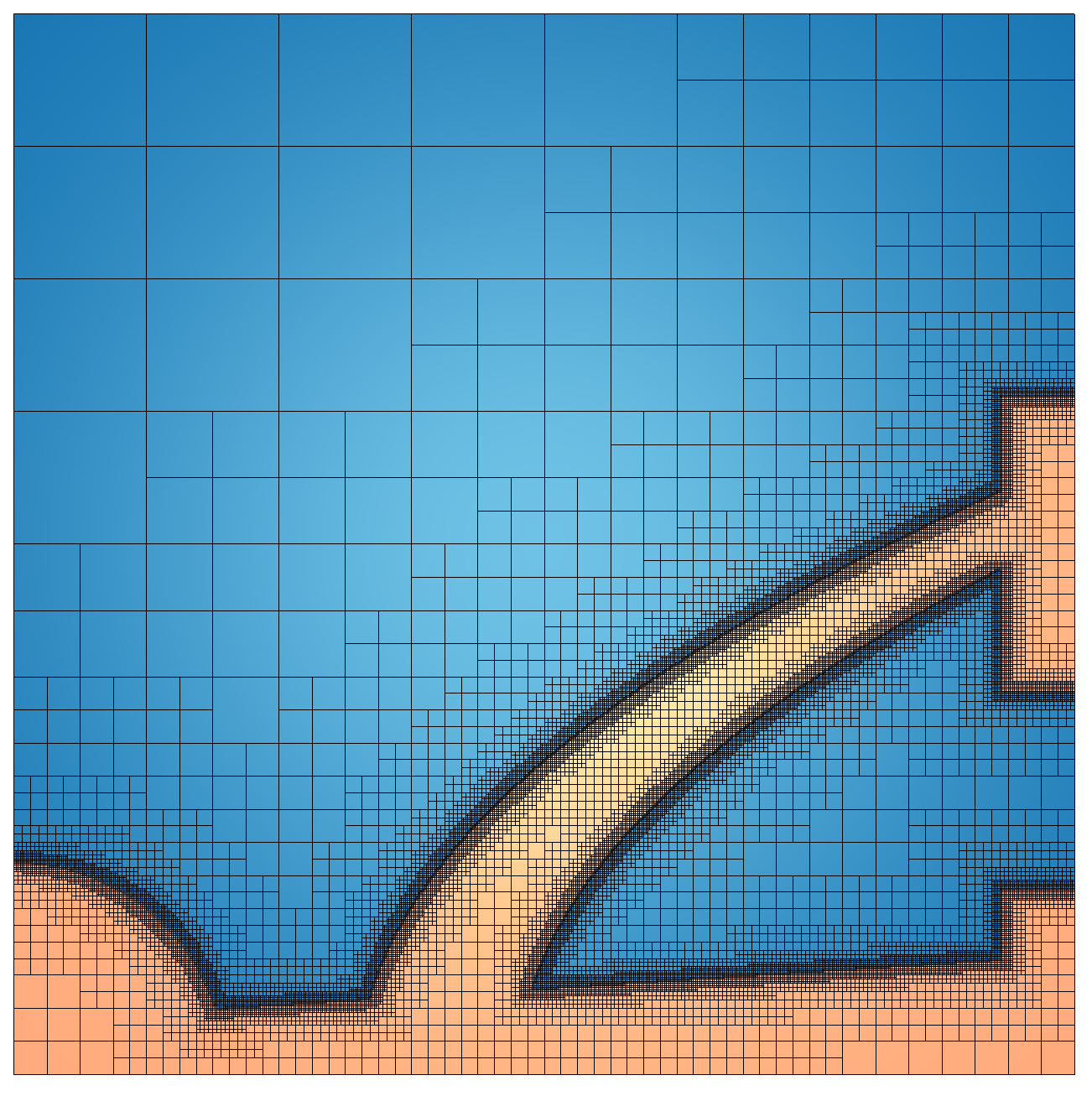} \hspace{2mm} &
   \includegraphics[height=0.3\textwidth]{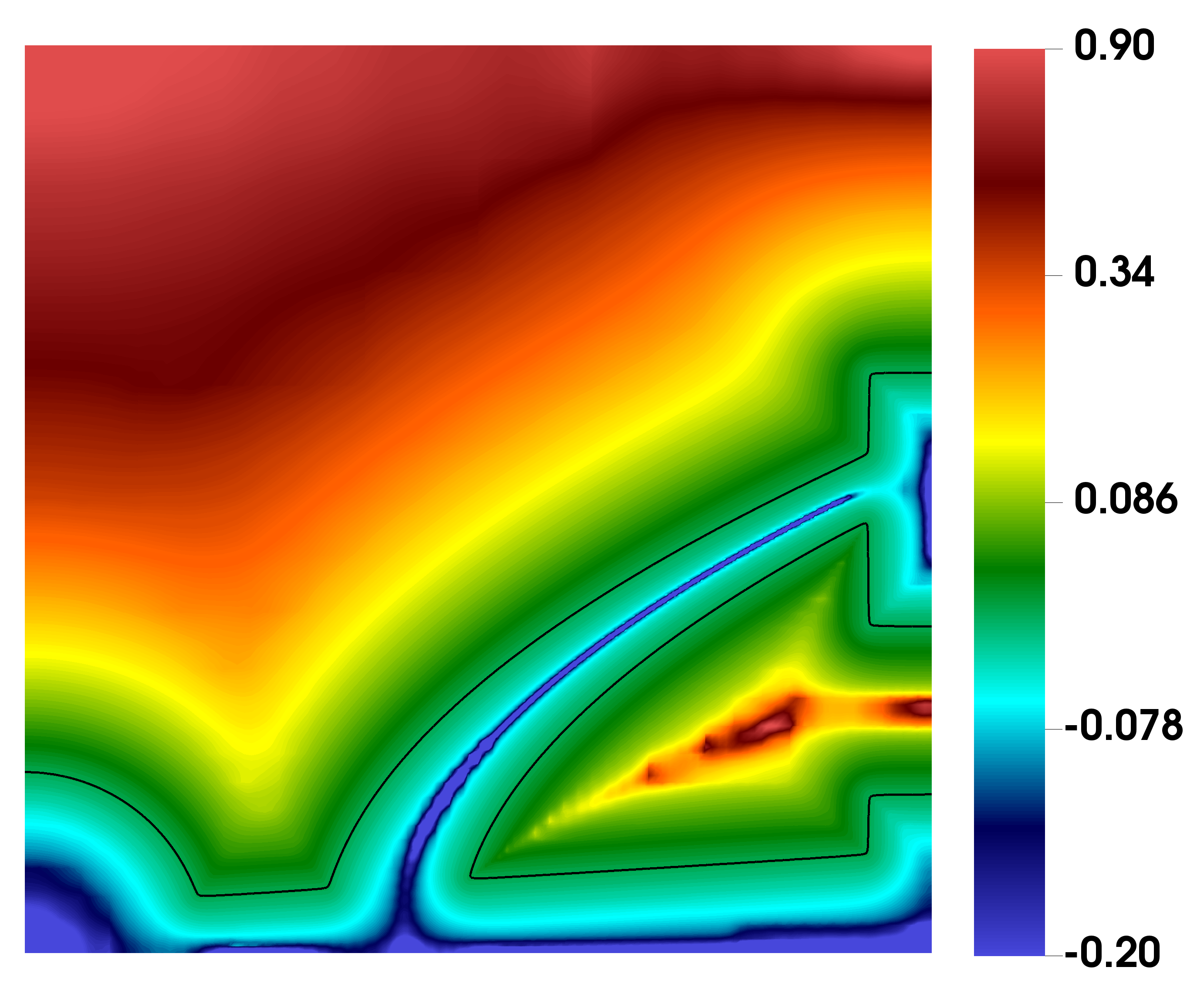} \hspace{2mm} \vspace{-2mm}\\
\textrm{(a)} &
\textrm{(b)} &
\textrm{(c)} \\
\end{array}$
\end{center}
\vspace{-7mm}
\caption{Multimaterial domain for a target application.
(a) Domain modeled using geometric primitives with $\mathcal{G}(\bx_B)= 1$ (orange) and $-1$ (blue), (b)
background mesh with adaptive mesh refinements around the zero level set of
$\mathcal{G}(\bx_B)$, and (c) distance function calculated on the background mesh \cgreen{\cite{belyaev2015variational}}
to be used as $\sigma(x)$ in \eqref{eq_F_full_sigma}.}
\label{fig_nrd_ls}
\end{figure}

While surfaces such as the circular interface in Fig. \ref{fig_square_circle} and sinusoidal boundary in Fig. \ref{fig_boundary} are straightforward to define using smooth
level-set functions, more intricate domains are typically defined
as a combination of non-smooth functions for different geometric primitives, which are not well suited for
our penalization-based formulation \eqref{eq_F_full_sigma}.
Consider for example a two-material application problem in
Fig. \ref{fig_nrd_ls}(a) where the domain is prescribed as a combination of geometric primitives for
a circle, rectangle, parabola, and a trapezium.
The resulting step function $\mathcal{G}(x_B)$ is 1 inside one material and -1 inside the other material.
For such cases, we start with a coarse
background mesh and use adaptive mesh refinement
around the zero level set of $\mathcal{G}(x_B)$, see Fig \ref{fig_nrd_ls}(b).
Then we compute a discrete distance function \cgreen{using the
p-Laplacian solver of \cite{belyaev2015variational}, Section 7,}
from the zero level set of $\mathcal{G}(x_B)$, see Fig \ref{fig_nrd_ls}(c).
The advantage of using this distance function in \eqref{eq_F_full_sigma} is that
it is (i) generally smoother and (ii) maintains the location of the zero level set.

\subsection{Setting $\mathcal{S}$ for Interface Fitting}
\label{sec_marking}

\cgreen{For interface fitting, the set $\mathcal{S}$ contains the nodes that
are used to discretize the material interface in the mesh.
As demonstrated in this section, the accuracy of the fitting to the
target surface using \eqref{eq_F_full_sigma} depends on the combination of the
(i) mesh topology around the interface,
and (ii) the shape of the target surface.
Thus, for a given initial mesh and implicit interface,
it is important to choose the fictious material indicator of elements such
that the resulting material interface is compatible with the target surface.}

\begin{figure}[bt!]
\begin{center}
$\begin{array}{cccc}
\includegraphics[height=0.25\textwidth]{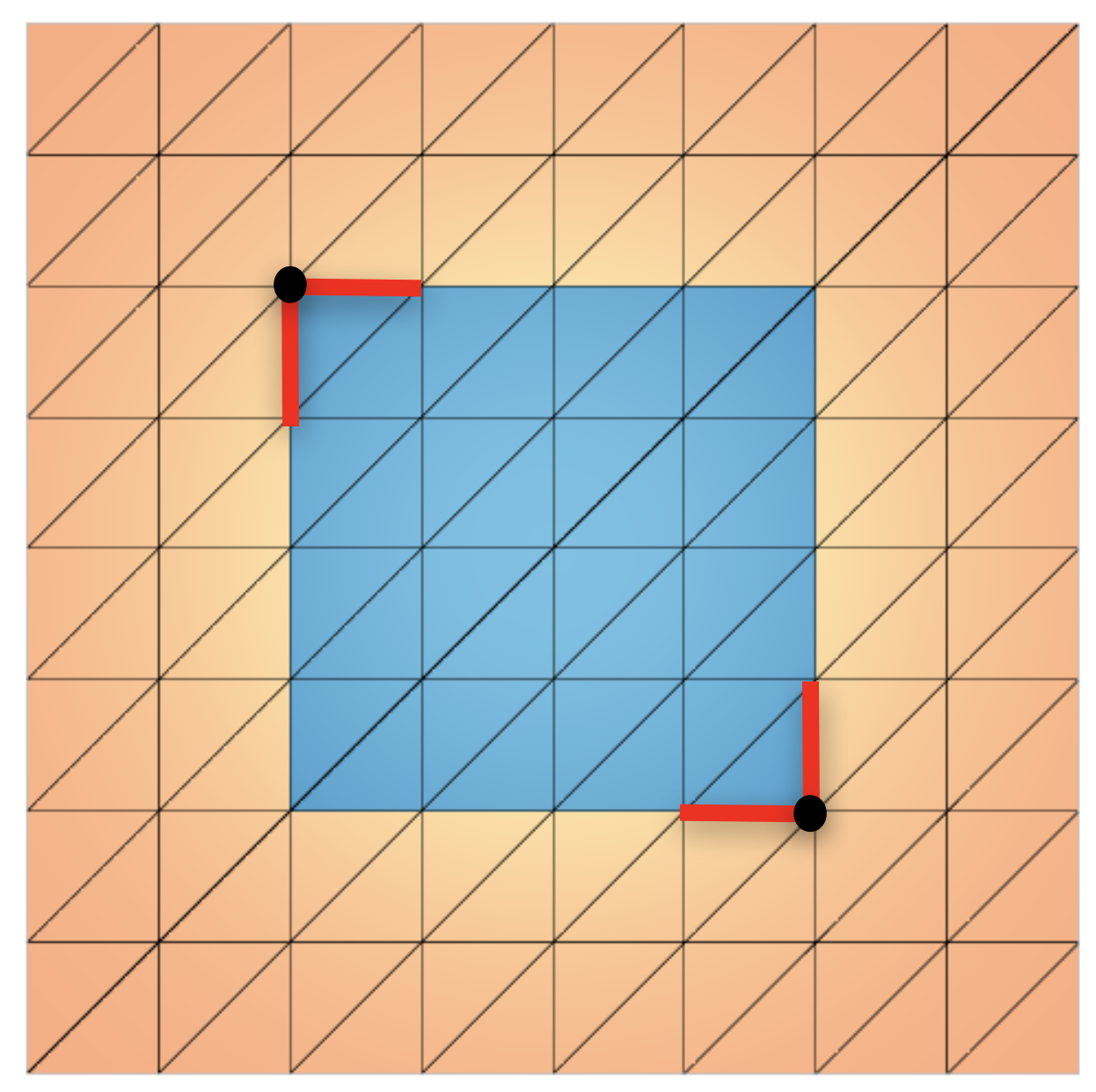} &
\includegraphics[height=0.25\textwidth]{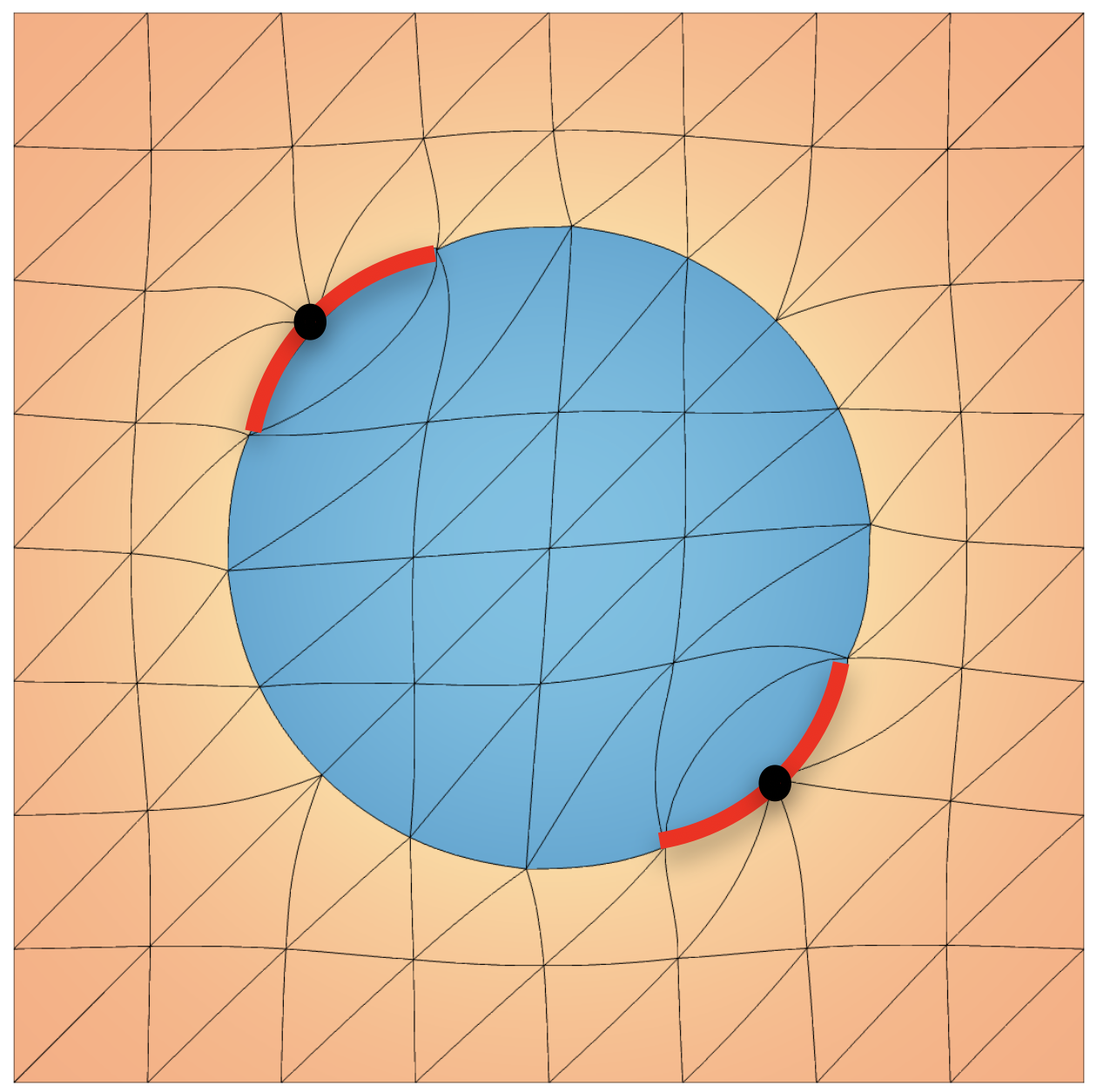} &
\includegraphics[height=0.25\textwidth]{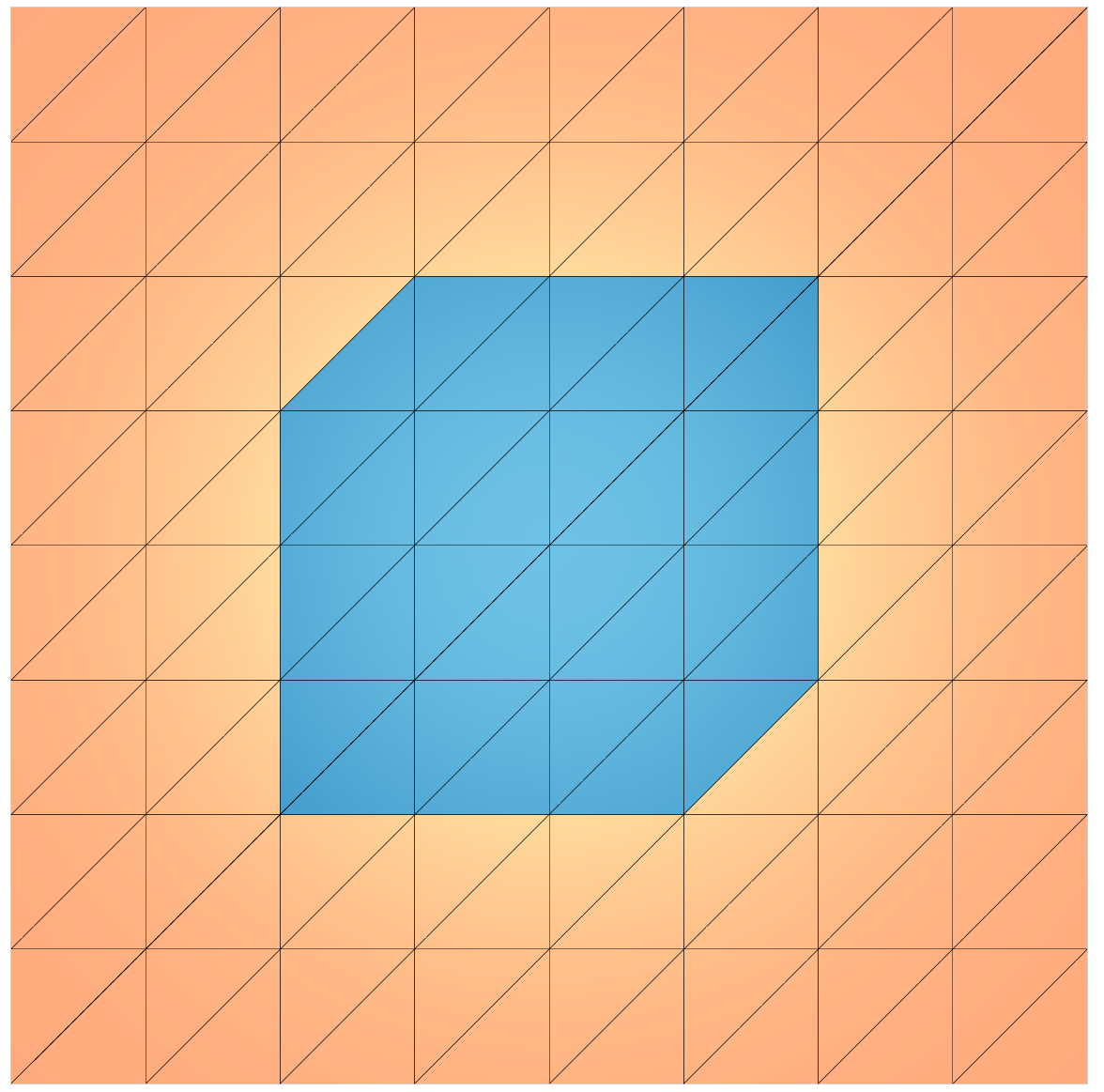} &
\includegraphics[height=0.25\textwidth]{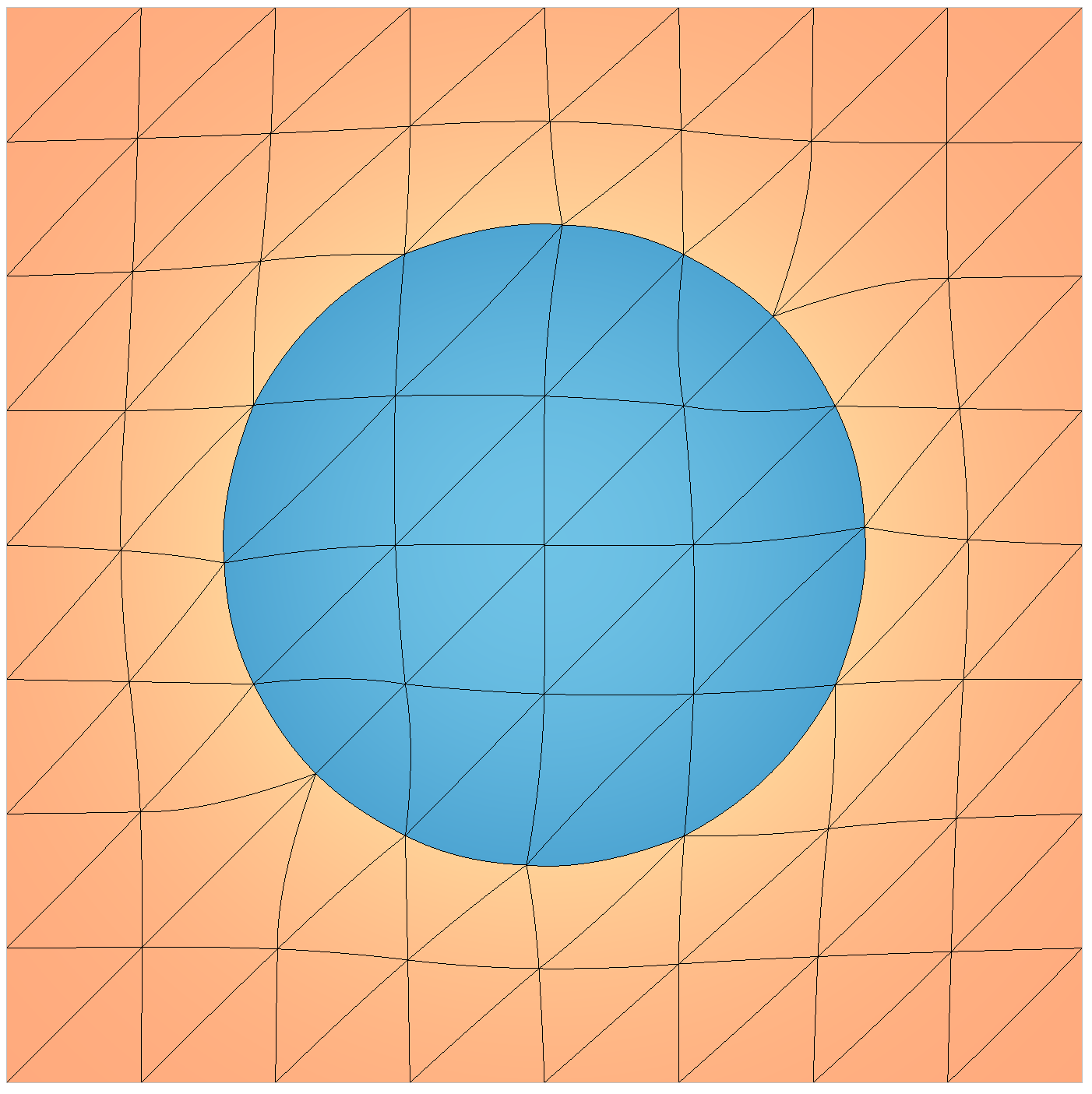} \vspace{-2mm}\\
\textrm{(a)} & \textrm{(b)} & \textrm{(c)} & \textrm{(d)} \\
\end{array}$
\end{center}
\vspace{-7mm}
\caption{(a) Triangular mesh with material indicators of elements using \eqref{eq_eta} and the (b) optimized mesh. (c) Mesh with material indicators of elements using \eqref{eq_eta2} and the resulting (d) optimized mesh.}
\label{fig_square_circle_tri}
\end{figure}

Consider for example the triangular mesh shown in Figure \ref{fig_square_circle_tri}(a) which must be aligned to a circular level set. \cblue{A naive approach for setting the material indicators is to partition the mesh into two fictitious materials
based on the level-set function sampled at the set $\mathcal{Q}$ of quadrature points inside each element. For example, the material indicator $\eta_E$ for element $E$ can be set using the integral of $\sigma(x)$ as}:
\begin{equation}
\label{eq_eta}
\eta_E=
\begin{cases}
    0,& \text{if } \int_E \sigma(x)\geq 0\\
    1,              & \text{otherwise}
\end{cases},
\end{equation}
\cblue{or using the maximum norm of $\sigma(x)$ as}:
\begin{equation}
\label{eq_eta_2}
\eta_E=
\begin{cases}
    0,& \text{if } \max_{q \in {\mathcal{Q}}} |\sigma(x_q)| \geq 0 \\
    1,              & \text{otherwise}.
\end{cases}.
\end{equation}
Using such approaches can lead to elements that have multiple adjacent faces as a part of the material interface of the mesh (highlighted in red). When the mesh deforms for fitting, this results in sub-optimal Jacobians in the elements at the vertex shared by the (highlighted) adjacent faces. This is evident in Figure \ref{fig_square_circle_tri}(b) where the material indicator is set using \eqref{eq_eta}.

The fundamental issue here is that whenever adjacent faces of an element are aligned to a level set,
the resulting mesh quality and fit can be sub-optimal depending on the complexity of the target shape/geometry.
To address this issue, we update the
material indicator for each element in the mesh as:
\begin{equation}
\label{eq_eta2}
\tilde{\eta}_E=
\begin{cases}
    \eta_E,& \text{if } N_{E,M} <= 1  \\
    1-\eta_E,& \text{else if } N_{E,M} = N_{E,F}-1 \\
    \eta_E, {\tt (optionally)\,mark\,E\,for\,splitting}  & \text{otherwise}
\end{cases},
\end{equation}
where $N_{E,M}$ is the number of faces of element $E$ that are part of the material interface,
and $N_{E,F}$ is the total number of faces of element $E$. Note that \eqref{eq_eta2}
is formulated as a two pass approach where we first loop over all elements with $\eta_E=0$
and then with $\eta_E=1$. This approach ensures that conflicting decisions are not
made for adjacent elements surrounding the interface.

\cblue{The first condition in \eqref{eq_eta2} keeps the original material indicator of an element when one of its faces is to be aligned to the level set. The second condition switches the material indicator of an element if all but one of the faces of an element is to be aligned, thus resulting in only one of its faces to be aligned after the switch. These two conditions are usually sufficient to guarantee that at most one face per element is set for alignment in triangular meshes.}
Figures \ref{fig_square_circle_tri}(c, d) show an example of material interface that results from \eqref{eq_eta2},
which avoids the issue of sub-optimal Jacobian at the vertex shared by
adjacent faces.

\begin{figure}[tb!]
\begin{center}
$\begin{array}{cccc}
\includegraphics[height=0.25\textwidth]{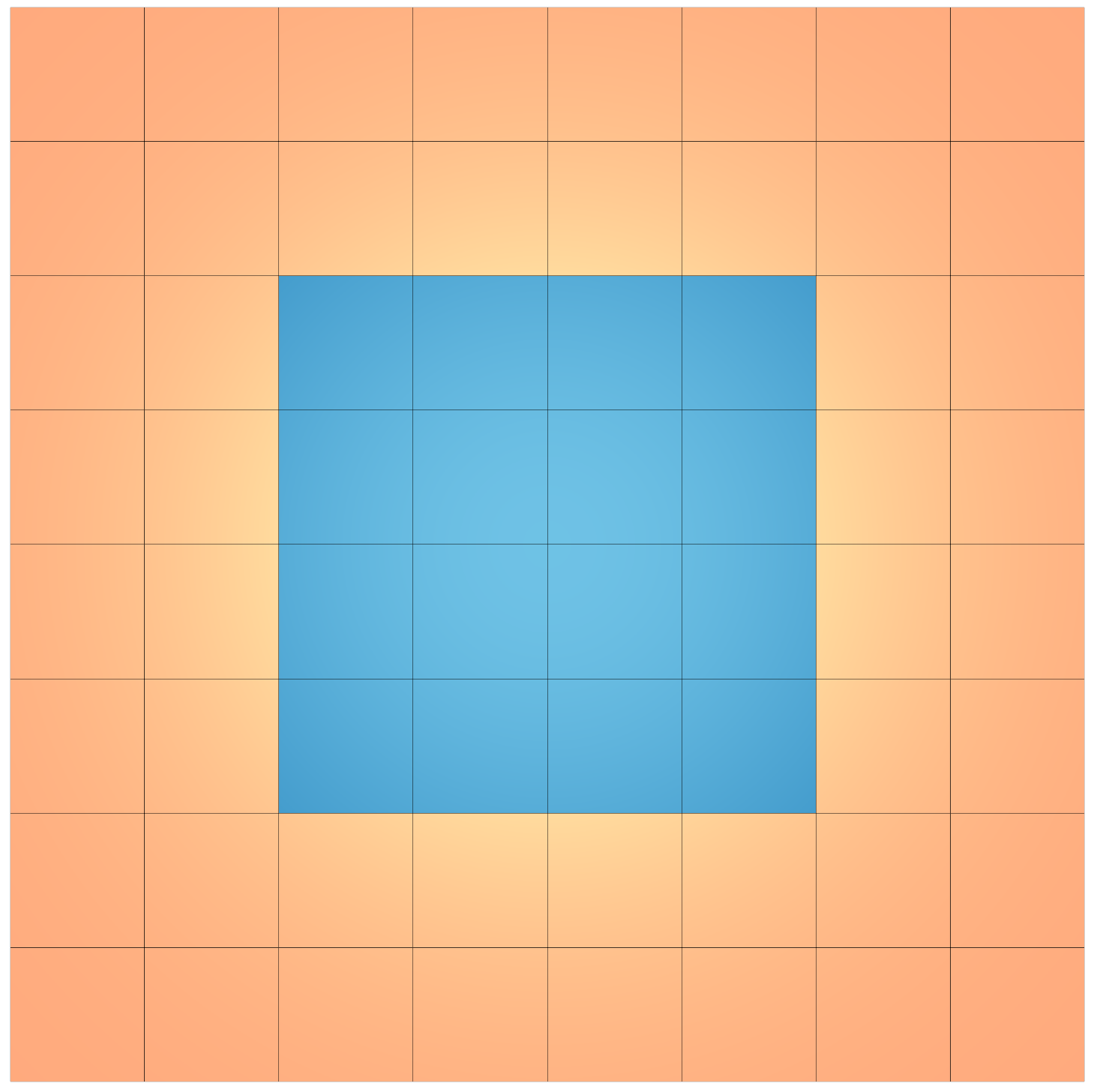} &
\includegraphics[height=0.25\textwidth]{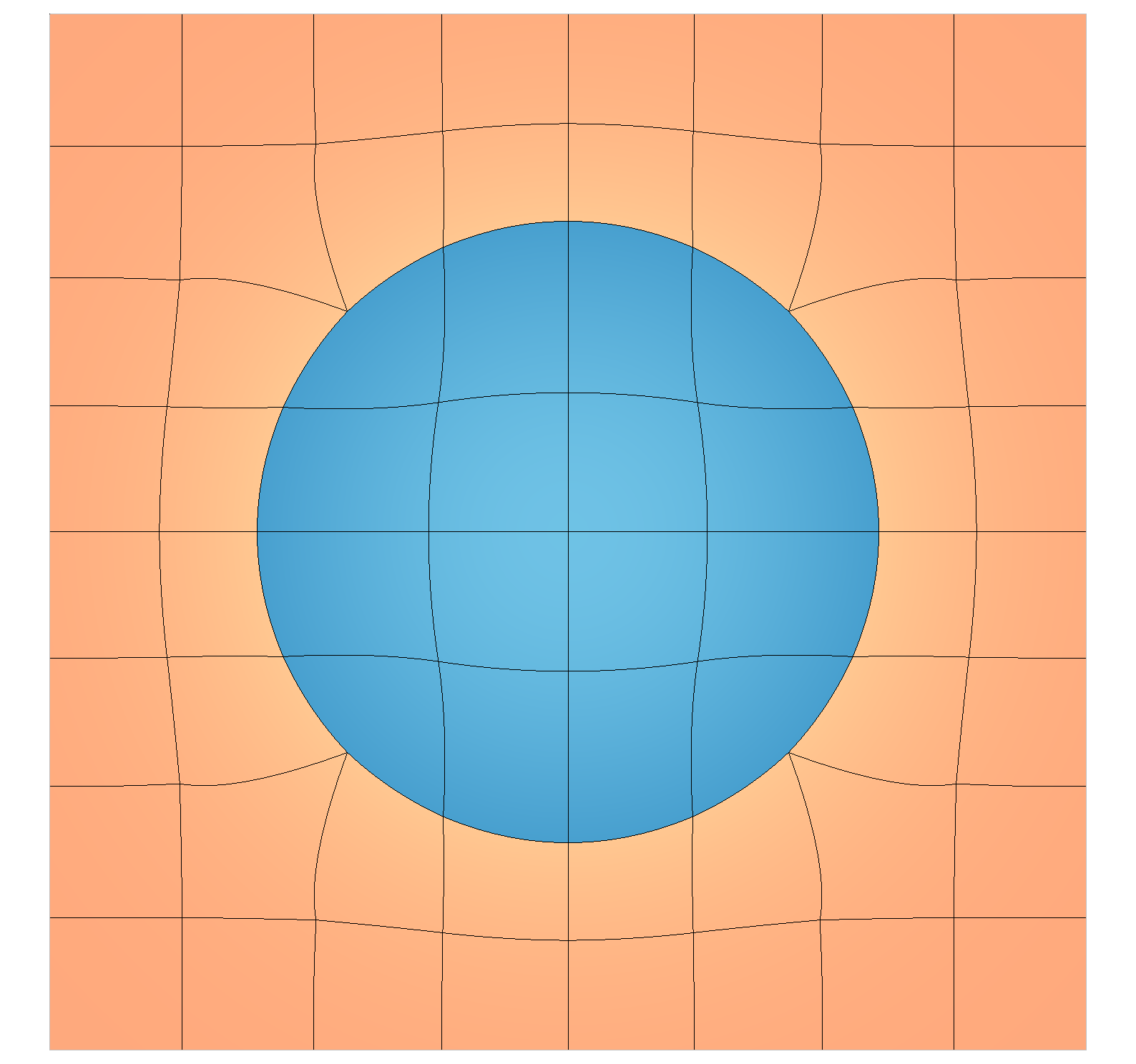} &
\includegraphics[height=0.25\textwidth]{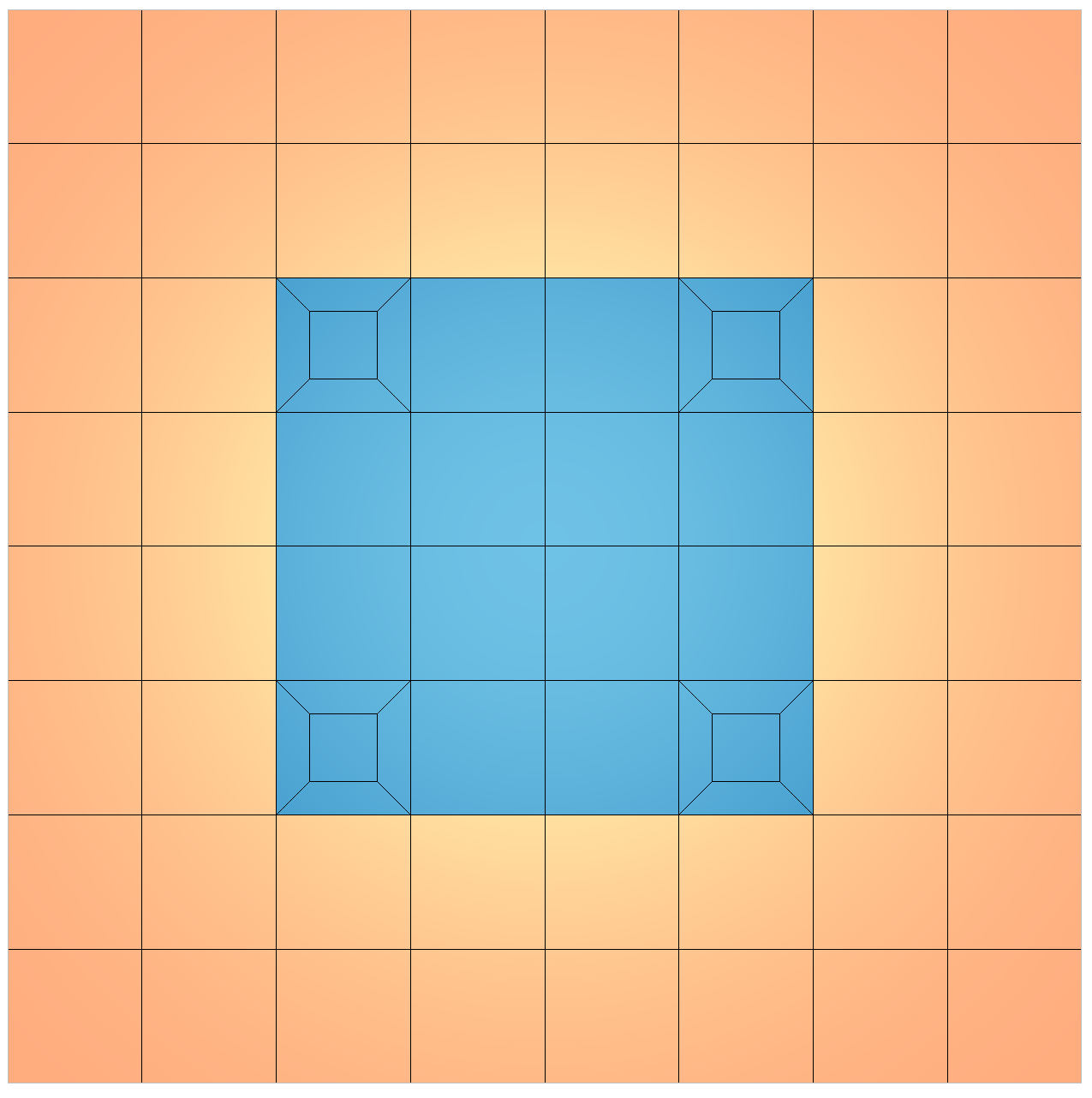} &
\includegraphics[height=0.25\textwidth]{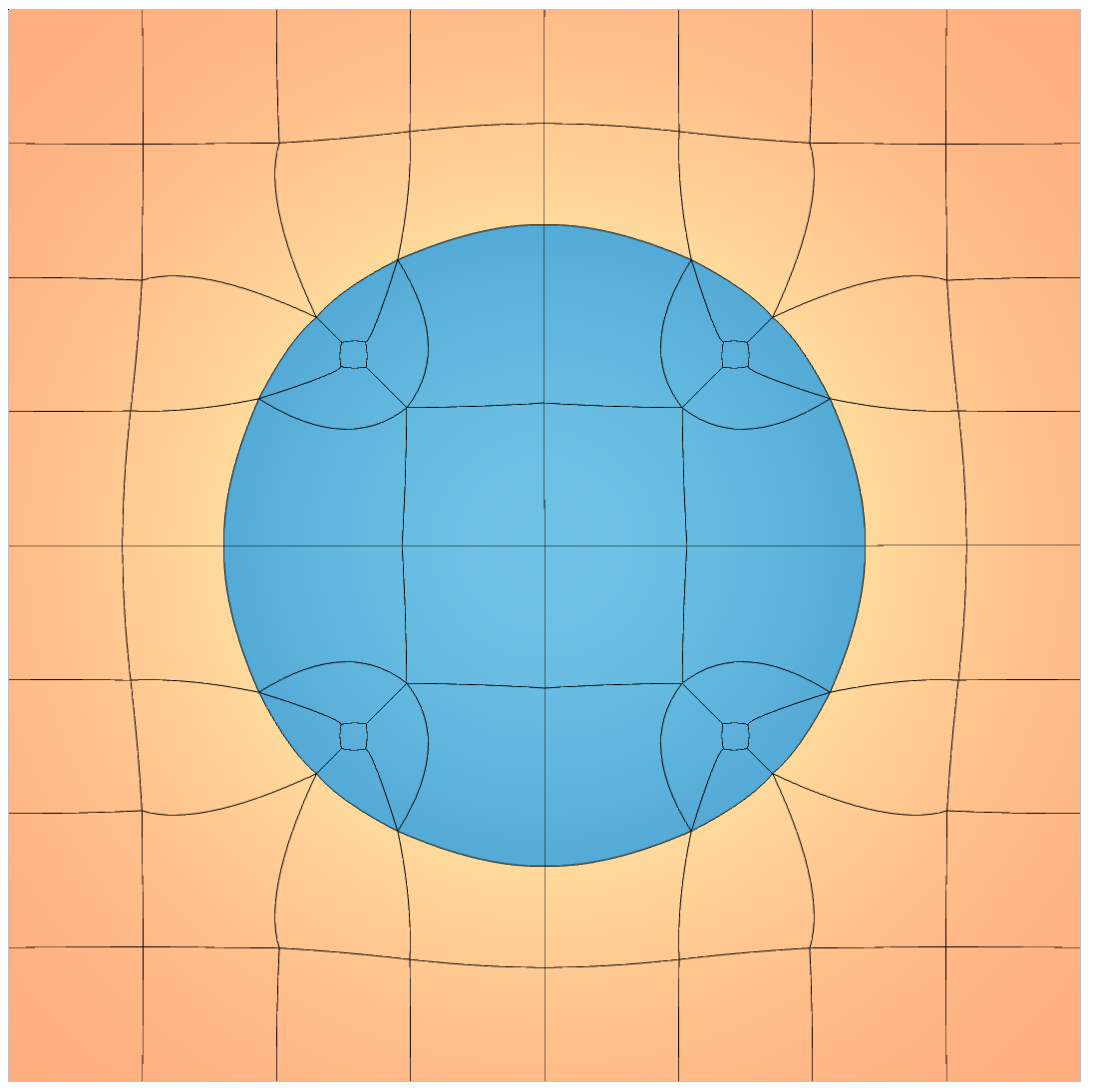} \vspace{-2mm} \\
\textrm{(a)} & \textrm{(b)} & \textrm{(c)} & \textrm{(d)} \\
\end{array}$
\end{center}
\vspace{-7mm}
\caption{(a) Quadrilateral mesh with material indicators using the strategy in \eqref{eq_eta} and the (b) optimized mesh. (c) Mesh material indicators set using
\eqref{eq_eta2} with conforming split introduced for elements with 2 faces marked
for fitting, and (d) the resulting optimized mesh.}
\label{fig_square_circle_quad}
\end{figure}

\cblue{In quadrilateral elements, when $1 < N_{E,M} < N_{E,F}-1$ (i.e. $N_{E,M}=2$), we can optionally
do conforming splits on each element to bisect the vertex connecting adjacent
faces that have been marked for fitting. This approach results in elements that have only 1 face marked for alignment, as long as the elements resulting from split keep the material indicator of the original element.
Conforming mesh refinements increase the computational cost due to increased number of degrees-of-freedom, but lead to a better mesh quality.
Figure \ref{fig_square_circle_quad}
shows an example of a comparison of a quadrilateral mesh fit to the circular level set using \eqref{eq_eta} and \eqref{eq_eta2}.}

\cred{Note that conforming splits can also be done for triangles, if needed, by connecting each of its vertex to the centroid of the triangle. For tetrahedra and hexahedra, conforming splits independent of adjacent elements are not yet possible, and we are currently exploring nonconforming refinement strategies.
Nonetheless, if such a situation arises where multiple adjacent faces of an element are marked for fitting, the proposed method will still align the mesh the best it can under the constraint of a prescribed threshold on minimum Jacobian in the mesh \eqref{eq_lsearch_detA}.}

\subsection{Adaptive Penalization Weight}
\label{sec_adaptive_sigma}

\begin{figure}[bt!]
\centerline{
   \includegraphics[height=0.5\textwidth]{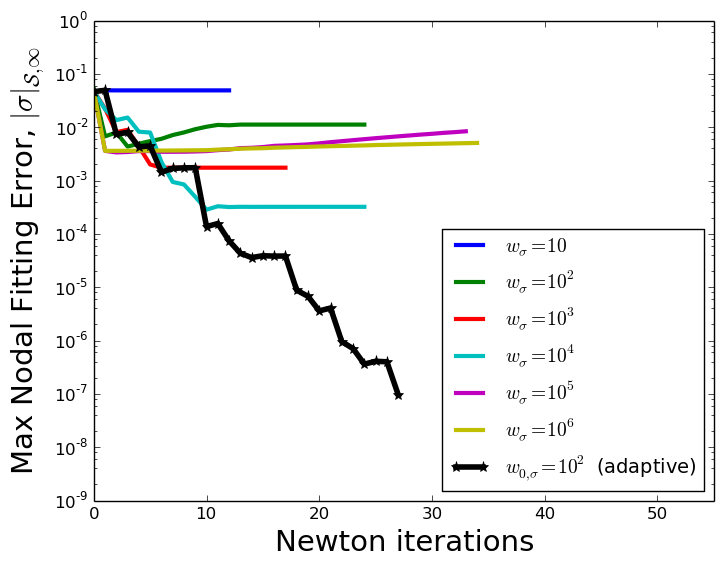}
   }
\caption{Impact of \ws\ on the surface fitting error.}
\label{fig_adapt_weight_sigma_circle}
\end{figure}

Recall that the balance between mesh quality and node fitting is controlled
by the penalization weight \ws\ in \eqref{eq_F_full_sigma}.
Numerical experiments show that use of a constant \ws\ requires tweaking on
a case-by-case basis, and can result in a sub-optimal fit if \ws\ is
too small, in which case the objective function is dominated by
the mesh quality metric term, or if \ws\ is too large, in which case the conditioning of the Hessian matrix is poor.
Figure \ref{fig_adapt_weight_sigma_circle} demonstrates how the maximum surface fitting error varies for a uniform quad mesh fit to a circular interface
(recall Figure \ref{fig_square_circle_quad}(a, b))
for different fixed values of \ws. Here, we define the maximum surface fitting error as \cblue{the maximum value of the level set function evaluated at the nodes $s \in \mathcal{S}$:}
\[
|\sigma|_{\mathcal{S},\infty} := \max_{s \in {\mathcal{S}}} |\sigma(x_s)|.
\]
Figure \ref{fig_adapt_weight_sigma_circle} shows that as we increase \ws\ from 1 to $10^{4}$, the fitting error decreases.
However, the error worsens if we increase $w_{\sigma}$ further.

To address this issue we use an adaptive approach for setting \ws\, where
we monitor $|\sigma|_{\mathcal{S},\infty}$, and scale \ws\
by a user-defined constant ($\alpha_{\sigma}=10$ by default) if the relative decrease in the
maximum nodal fitting error between subsequent Newton iterations is below a
prescribed threshold ($\epsilon_{\Delta \sigma}=0.001$ by default). That is,
\begin{equation}
\label{eq_w_adaptive}
 w_{k+1,\sigma} =
 \begin{cases}
  \alpha_{\sigma}\cdot w_{k,\sigma} & \text{ if $\frac{|\sigma|_{k,\mathcal{S},\infty}-|\sigma|_{k+1,\mathcal{S},\infty}}{|\sigma|_{k+1,\mathcal{S},\infty}} < \epsilon_{\Delta \sigma}$} \\
  w_{k,\sigma} & \text{ otherwise }
\end{cases},
\end{equation}
where we use the subscript $k$ to denote a \cblue{quantity} at the $k$th Newton iteration.
Figure \ref{fig_adapt_weight_sigma_circle}
shows that this adaptive approach for setting \ws\ significantly
improves the quality of the mesh fit to the desired level set.
Updating the value of \ws\ changes the definition of the objective function
\eqref{eq_F_full_sigma}, which requires some modifications of the line search
and convergence criterion of the Newton iterations to
achieve overall convergence, compared to the constant \ws\ case;
details will follow in Section \ref{sec_converge}.
Nevertheless, our numerical tests suggest that this impact is negligent
in comparison to the improvement of the fitting error.

\subsection{Convergence \& Line-Search Criterion}
\label{sec_converge}

Recall that in the general TMOP approach, the line-search and convergence
criteria for the Newton's method are based on the magnitude and the
derivatives of the objective function, see Section \ref{sec_ra}.
In the penalization-based formulation \eqref{eq_F_full_sigma},
the current criteria do not suffice because
the magnitude and derivatives of the objective function depend on
the penalization weight $w_{k,\sigma}$, which can change between
subsequent Newton iterations due to \eqref{eq_w_adaptive}.

We modify our line-search criteria by adding two additional inequalities,
namely, $\alpha$ in \eqref{eq_r_adaptivity_solve} is chosen to ensure:
\begin{equation}
\label{eq_lsearch_sigma}
|\sigma|_{k+1,\mathcal{S},\infty} < 1.2~|\sigma|_{k,\mathcal{S},\infty},
\end{equation}
\begin{equation}
\label{eq_lsearch_detA}
{\tt min}\big(\det(A(\bx_{k+1}))\big) > 0.001 \cdot {\tt min}\big(\det(A(\bx_0))\big).
\end{equation}
The inequality \eqref{eq_lsearch_sigma} prevents sudden
jumps of the fitting error, \cgreen{and the scaling factor 1.2 has been chosen empirically}.
The constraint \eqref{eq_lsearch_detA} is mostly applicable in regimes when
\ws\ is big enough to make the quality term $F_{\mu}$ effectively inactive.
Such regimes represent a situation when one is willing to sacrifice mesh
quality for more accurate fitting.
When the quality term $F_{\mu}$ is relatively small, it may be unable to
prevent the appearance of infinitesimally small positive Jacobians;
the constraint \eqref{eq_lsearch_detA} is used to alleviate this situation.

The convergence criterion is also modified to utilize the fitting error, i.e.,
the Newton's method is used until $|\sigma|_{\mathcal{S},\infty}$ is
below a certain user-specified threshold ($\epsilon_{\sigma}=10^{-5}$ by default). We also use an optional convergence criterion based on a limit on the maximum number of consecutive Newton iterations through which the penalization weight \ws\ is adapted using \eqref{eq_w_adaptive}; this limit is $N_{\sigma}=10$ by default. This latter criterion avoids excessive computations in cases where the mesh topology does not allow the fitting error to reduce beyond a certain limit.

\subsection{Derivatives}
\label{sec_derivatives}

As our default choice for nonlinear optimization is the Newton's method, we
must compute first and second order derivatives of $F_{\mu}$ and $F_{\sigma}$ with respect to the mesh nodes.
\cgreen{The definition of the derivatives of $F_{\mu}$ is given in
Sections 3.3 and 3.4 of \cite{camier2022accelerating},
and here we focus on the derivatives of $F_{\sigma}$.}
Let the FE position function be $x = (x_1 \dots x_d)^T$ where $d$ is the
space dimension; each component can be written as
$x_a(\bar{x}) = \sum_i x_{a,i} \bw_i(\bar{x})$, where $x=x(\bar{x})$,
see Section \ref{sec_mesh}.
Then the Newton's method solves for the full vector
\[
\bx = (x_{1,1} \dots x_{1,N_x}, x_{2,1} \dots x_{2,N_x} \dots x_{d,N_x})^T
\]
that contains the positions of all mesh nodes.
The formulas for the first and second derivatives of $F_{\sigma}$
are the following:
\begin{equation}
\label{eq_derivatives}
\begin{split}
\frac{\partial F_{\sigma}(\bx)}{\partial x_{a,i}}
&= 2 \omega_{\sigma} \sum_{s \in \mathcal{S}}
  \sigma(x_s) \frac{\partial \sigma(x_s)}{\partial x_{a}}
  \frac{\partial x_a(\bar{x}_s)}{\partial x_{a,i}}
= 2 \omega_{\sigma} \sum_{s \in \mathcal{S}}
  \sigma(x_s) \frac{\partial \sigma(x_s)}{\partial x_{a}}
  \bw_i(\bar{x}_s), \\
\frac{\partial^2 F_{\sigma}(\bx)}{\partial x_{b,j} \partial x_{a,i}}
&= 2 \omega_{\sigma} \sum_{s \in \mathcal{S}}
  \left(
  \frac{\partial \sigma(x_s)}{\partial x_{b}}
  \frac{\partial \sigma(x_s)}{\partial x_{a}} +
  2 \omega_{\sigma} \frac{\partial^2 \sigma(x_s)}{\partial x_{b} \partial x_{a}}
  \right) \bw_i(\bar{x}_s) \bw_j(\bar{x}_s), \\
a,b &= 1 \dots d, \quad i,j = 1 \dots N_x.
\end{split}
\end{equation}
The above formulas require the spatial gradients of $\sigma$ at the current
positions $\{x_s\}_{s \in \mathcal S}$ of the marked nodes.
These gradients can be closed-form expressions, when $\sigma$ is prescribed
analytically, or the gradients are obtained from the background mesh
$\mathcal{M}_B$ (see Section \ref{sec_level_set_rep}),
when $\sigma$ is a discrete function.

\subsection{Algorithm/Summary}
\label{sec_algo}

In this section, we summarize our penalization-based method for boundary and
interface fitting with TMOP.
The inputs to our method are the active/current mesh $\mathcal{M}$ that is
defined through the global vector $\bx$ of nodal positions,
a user-selected target construction option to form $W$ as in \eqref{eq_W},
a mesh quality metric $\mu$,
a source/background mesh with nodal coordinates $\bx_B$ along with the level set
function $\sigma(x_B)$ as explained in Section \ref{sec_level_set_rep},
initial penalization weight \ws,
parameters for adaptive penalization-weight ($\alpha_{\sigma}$ and
$\epsilon_{\Delta \sigma}$ as in Section \ref{sec_adaptive_sigma}),
and parameters for convergence criterion $\epsilon_{\sigma}$ and
$N_{\sigma}$ as in Section \ref{sec_converge}.
Algorithm \ref{algo_fit} summarizes our penalization-based method where we
use subscript $k=0\dots N_{opt}$ to denote different \cblue{quantities} at the
$k$th Newton iteration.

\begin{algorithm}[h!]
\label{algo_fit}
\SetAlgoLined
\KwIn{$\bx$, $\mu$, $\bx_B$, $\sigma(x_B)$, $\alpha_{\sigma}$, $\epsilon_{\Delta \sigma}$, $w_{\sigma}$, $\epsilon_{\sigma}$, $N_{opt}$}
\KwOut{$\bx_s$}
$n_{\sigma}:=0$,\,\,$k=0$\,\,,$\bx_{0}=\bx$\\
Determine $\mathcal{S}$, the set of nodes for fitting (Section \ref{sec_marking})\\
$\sigma(\bx_0) = \mathcal{I}(\bx_0, \bx_B, \sigma(\bx_B))$\\
\While{$|\sigma|_{k,\mathcal{S},\infty}>\epsilon_{\sigma}$ \&\& $n_{\sigma} < N_{\sigma}$ \&\& $k < N_{opt}$}{
    $W_i = I$ for each quadrature point $i$ ~\cite{knupp2019target}.\\
    \cblue{$\mathcal{H}(\bx_k) \bdx = \mathcal{J}(\bx_k)$ $\rightarrow$} solve using MINRES (Section \ref{sec_ra}). \\
    $\bx_{k+1} = \bx_{k} - \alpha \bdx$, with $\alpha$ determined using line-search (Section \ref{sec_converge}). \\
    \eIf{$\frac{|\sigma|_{k,\mathcal{S},\infty}-|\sigma|_{k+1,\mathcal{S},\infty}}{|\sigma|_{k+1,\mathcal{S},\infty}} < \epsilon_{\Delta \sigma}$} {
       $w_{k+1,\sigma} \rightarrow \alpha_{\sigma}\cdot w_{k,\sigma}$ \\
       $n_{\sigma}=0$
    }
    {
       $n_{\sigma} = n_{\sigma}+1$
    }
    $\sigma(\bx_{k+1}) = \mathcal{I}(\bx_{k+1}, \bx_B, \sigma(\bx_B))$\\
    $k = k+1$\\
}
$\bx = \bx_{k}$ \label{algo_end}
\caption{Implicit Meshing}
\end{algorithm}

\section{Numerical Results}
\label{sec_results}

In this section, we demonstrate the main properties of the method
using several examples. The presented tests use $W=I$ as the target matrix, and the following shape metrics:
\begin{equation}
\begin{split}
  \color{red}\mu_2     &= \frac{|T|^2}{2\tau} - 1, \\
  \color{red}\mu_{303} &= \frac{|T|^2}{3\tau^{2/3}} - 1,
\end{split}
\end{equation}
where $|T|$ and $\tau$ are the Frobenius norm and determinant of $T$, respectively.
Both metrics are polyconvex in the sense of \cite{Garanzha2010, Garanzha2014},
i.e., the metric integral $F_{\mu}$ in \eqref{eq_F_full} theoretically has a minimizer.
Exploring the effect of the \cblue{smoothness of $\sigma(x)$} on the convexity of the objective function \eqref{eq_F_full_sigma}
will be the subject of future studies.

Our implementation utilizes MFEM, a finite element library that supports arbitrarily high-order spaces and meshes \cite{MFEM2020}. This implementation is freely available at \url{https://mfem.org}.

\subsection{Fitting to a Spherical Interface}

\begin{figure}[b!]
\begin{center}
$\begin{array}{cccc}
\includegraphics[height=0.23\textwidth]{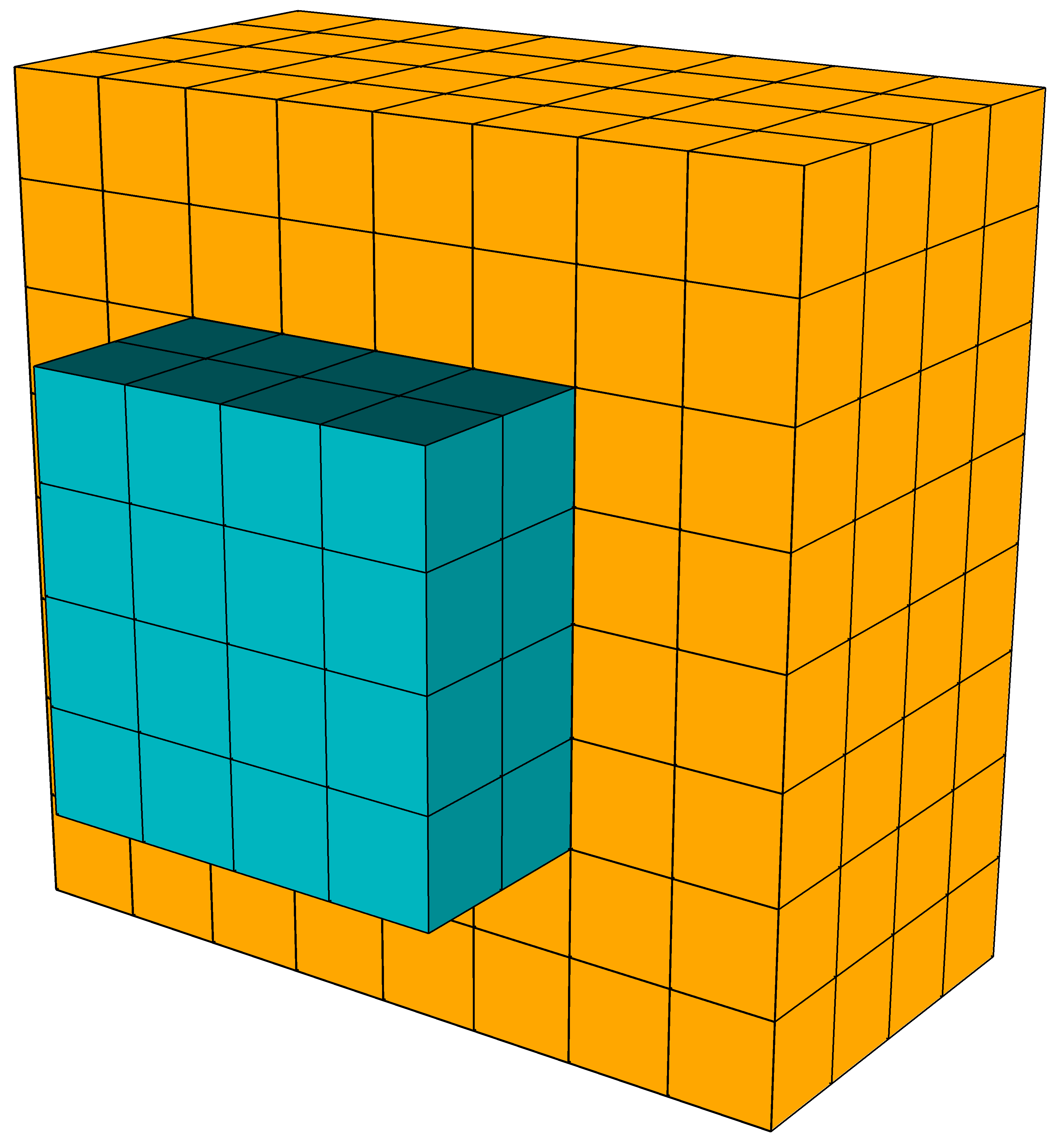} &
\includegraphics[height=0.23\textwidth]{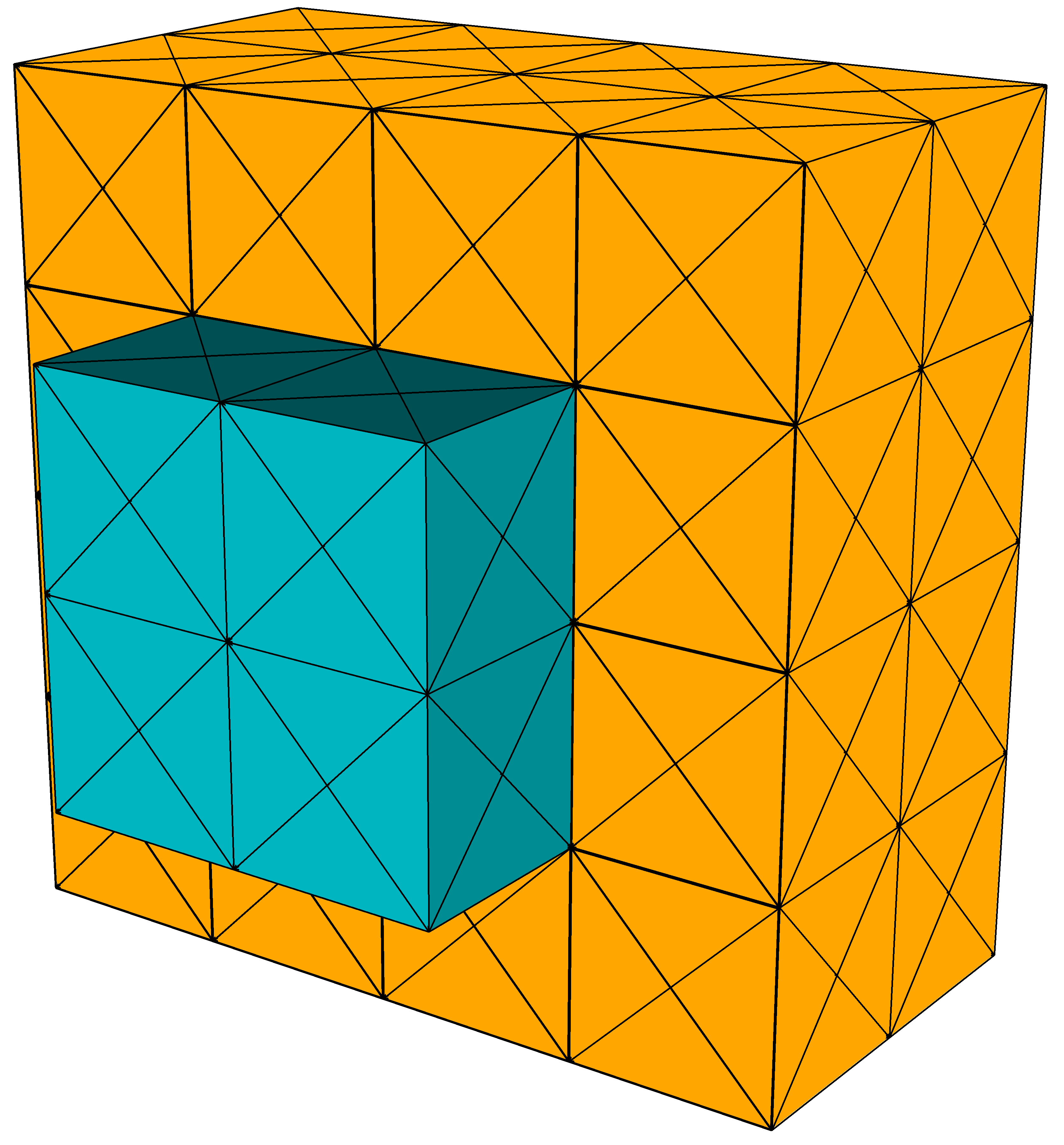} &
\includegraphics[height=0.22\textwidth]{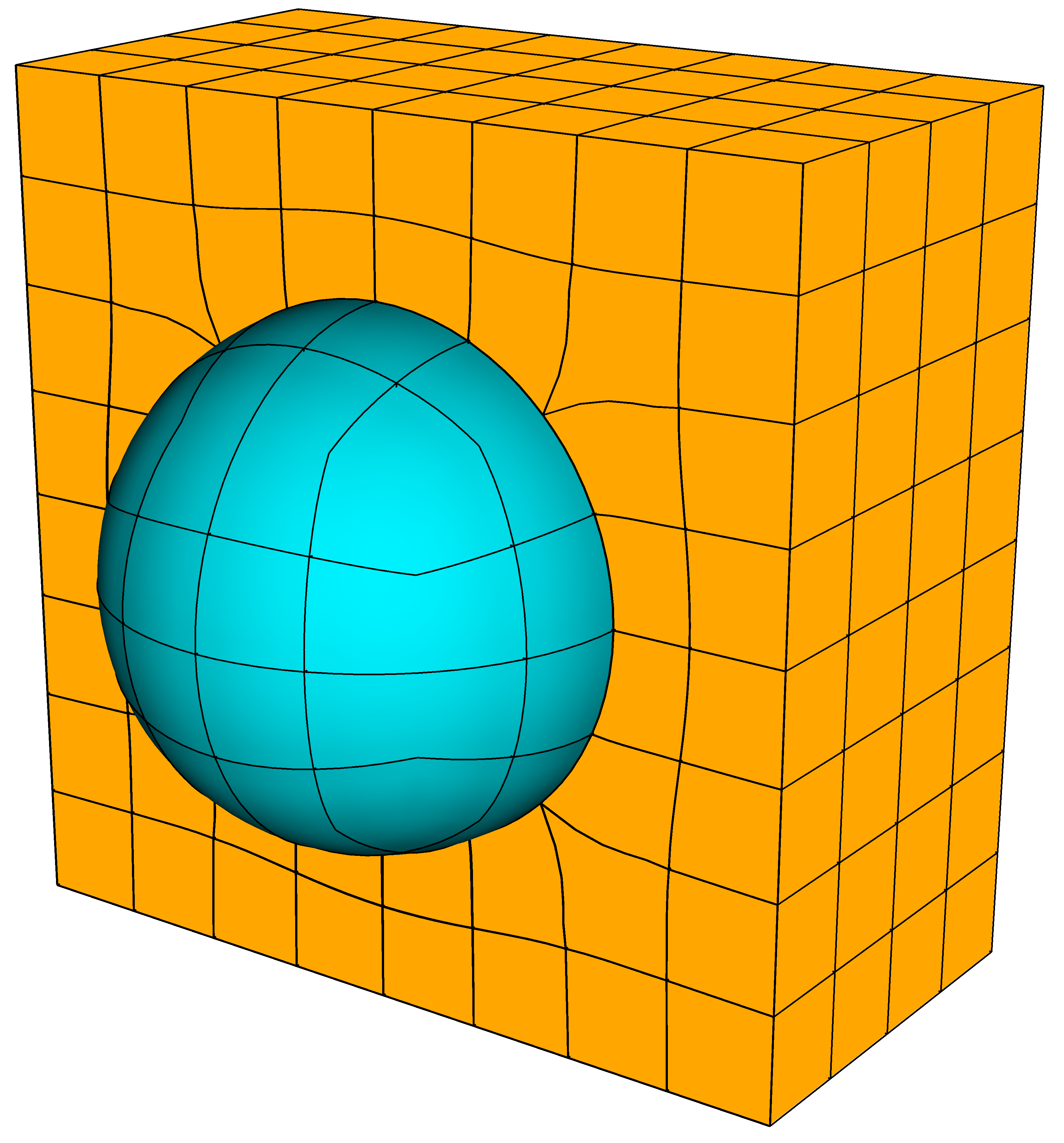} &
\includegraphics[height=0.22\textwidth]{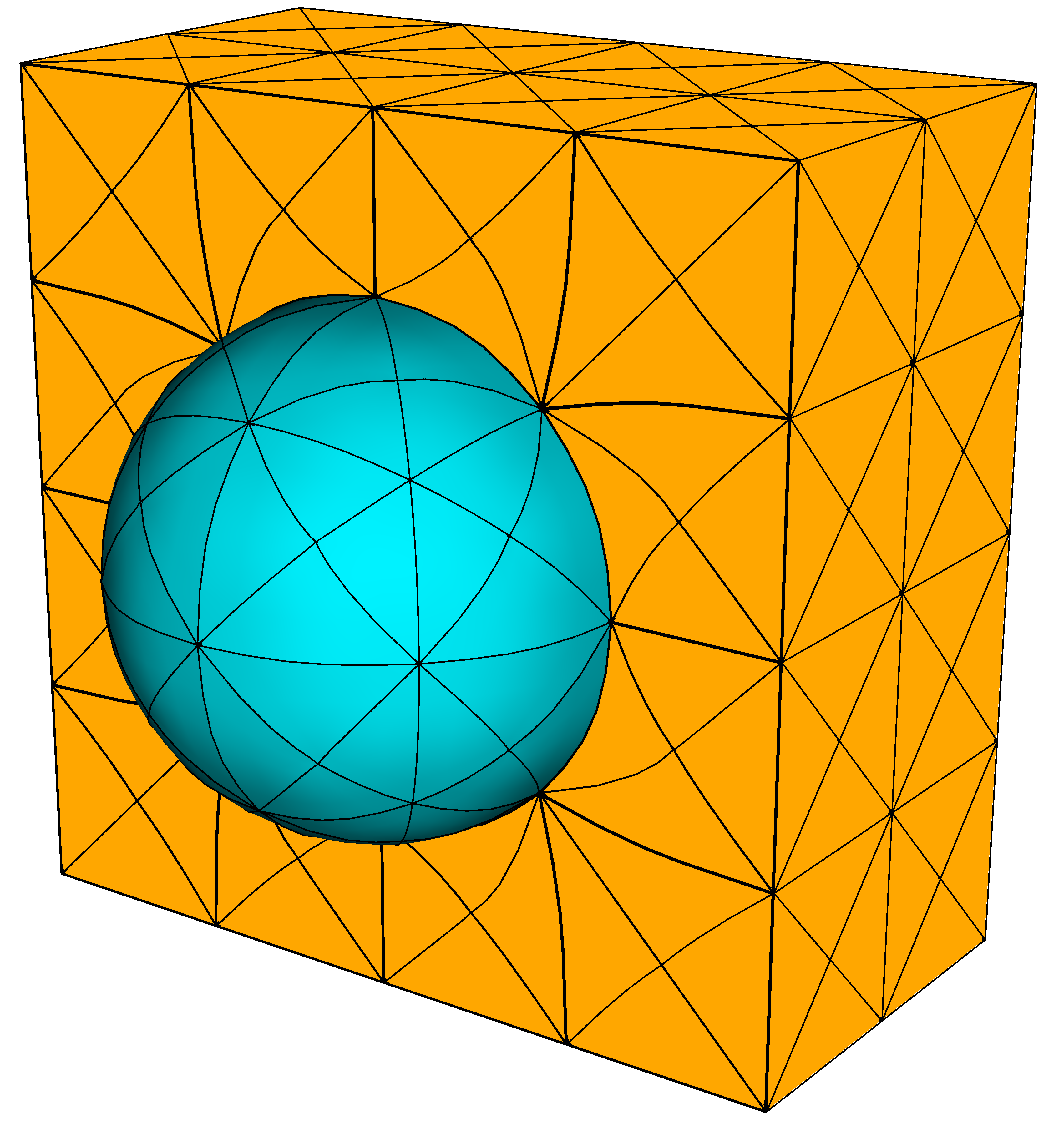} \vspace{-2mm}\\
\textrm{(a)} & \textrm{(b)} & \textrm{(c)} & \textrm{(d)} \\
\end{array}$
\end{center}
\vspace{-7mm}
\caption{Initial (a) hexahedral and (b) tetrahedral meshes showing the interfaces for the 3D surface fitting tests. Optimized (c) hexahedral and (d) tetrahedral mesh obtained using \eqref{eq_F_full_sigma}.}
\label{fig_3D_sphere_mesh}
\end{figure}

As a proof of concept, we adapt 3rd order hexahedral (hex) and tetrahedral (tet) meshes to align to a spherical surface; see Figure \ref{fig_3D_sphere_mesh}(a) and (b).
The domain is a unit-sized cube, $\Omega \in [0, 1]^3$, and
\cblue{the level set function $\sigma$ representing the sphere} is defined such that its zero isosurface is located at a distance of 0.3 from the center of the domain ($x_c = (0.5, 0.5, 0.5)$).
Although this level set is simple enough to be defined analytically,
the presented computations
represent and use $\sigma$ as a discrete finite element function.
Additionally, no background mesh is used in this example, i.e., $\bx_B = \bx$.

The initial hex mesh is an $8\times8\times8$ Cartesian-aligned mesh.
The material indicators are setup such that there are a total of 32 elements that have more than one face marked for fitting.
The tet mesh is obtained by taking a $4\times4\time4$ Cartesian-aligned hex mesh and splitting each hex into \cblue{24 tetrahedra sharing a vertex
at the center of the cube (4 tets-per-hex face)}.
The material indicators are setup such that all faces marked for fitting belong to different elements. The optimized meshes, shown in Fig. \ref{fig_3D_sphere_mesh}(c) and (d), have a maximum error of $O(10^{-10})$ at the interface with respect to the zero level set, which is achieved in 76 and 44 Newton iterations for the hexahedral and tetrahedral mesh, respectively. \cgreen{In the tetrahedral mesh, the minimum Jacobian decreases from $4\times 10^{-3}$ to $4.6\times 10^{-6}$ and in the hexahedral mesh, the minimum Jacobian decreases from $1.2\times 10^{-3}$ to $1.5\times 10^{-5}$ during mesh optimization as the mesh deforms to align to the target surface; in both cases the minimum Jacobian appears near the fitted faces. The decrease in the Jacobian at the interface is expected, with a lower bound on the Jacobian imposed by \eqref{eq_lsearch_detA} to ensure that the mesh stays valid for desired finite element computations.}
Note that the hexahedral mesh requires more iterations as compared to the tetrahedral mesh partly because there are multiple elements with adjacent faces marked for fitting, which requires more work from the adaptive weight mechanism to force those mesh faces to align with the spherical interface.

\subsection{Boundary and Interface Fitting for Geometric Primitive-Based Domains in 3D} \label{results_geom_prim_3D}
\begin{figure}[b!]
\begin{center}
$\begin{array}{c}
\includegraphics[height=0.35\textwidth]{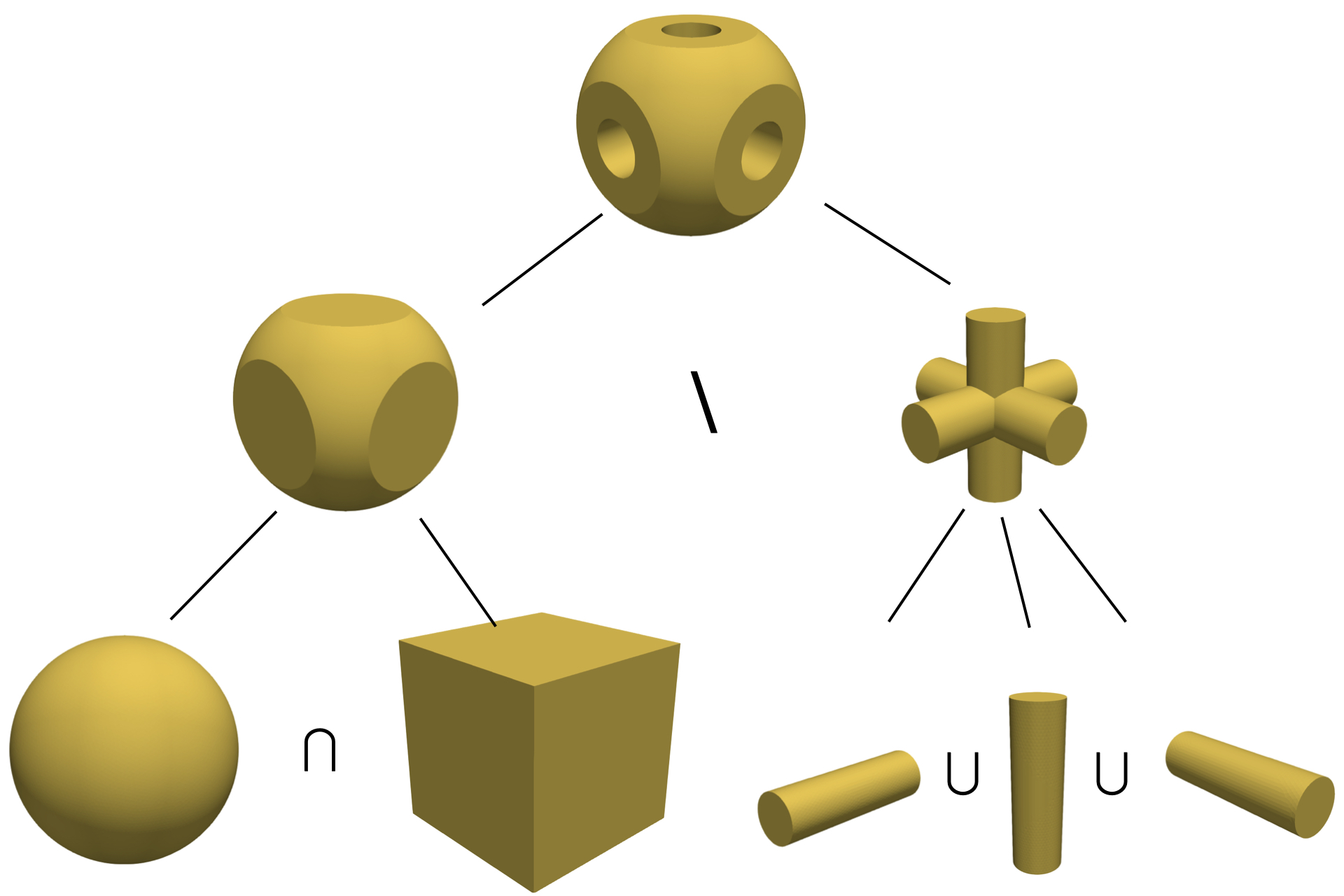}
\end{array}$
\end{center}
\vspace{-7mm}
\caption{CSG tree with geometric primitives used to define the target geometry. \cblue{Here,
$\bigcap$ denotes the geometric intersection operator, $\bigcup$ denotes the union operation, and $\backslash$ denotes the exclusion operator.}}
\label{fig_3D_geom_prim}
\end{figure}

\cred{This example demonstrates that the proposed fitting method can
be applied not only to internal interfaces, but also to high-order boundaries.}
The target domain is represented as a combination of simple
geometric primitives, namely, an intersection of a cube
(\emph{side}=0.5, centered around $x_c = (0.5, 0.5, 0.5)$) with a
sphere (\emph{radius}=0.3, centered around $x_c$) that has three Cartesian-aligned cylinders (\emph{radius}=0.15, \emph{length}=0.5) removed from it. Figure \ref{fig_3D_geom_prim} shows the CSG tree
with geometric primitives and the Boolean operations that are used to construct the target geometry.

\begin{figure}[tb!]
\begin{center}
$\begin{array}{cc}
\includegraphics[height=0.38\textwidth]{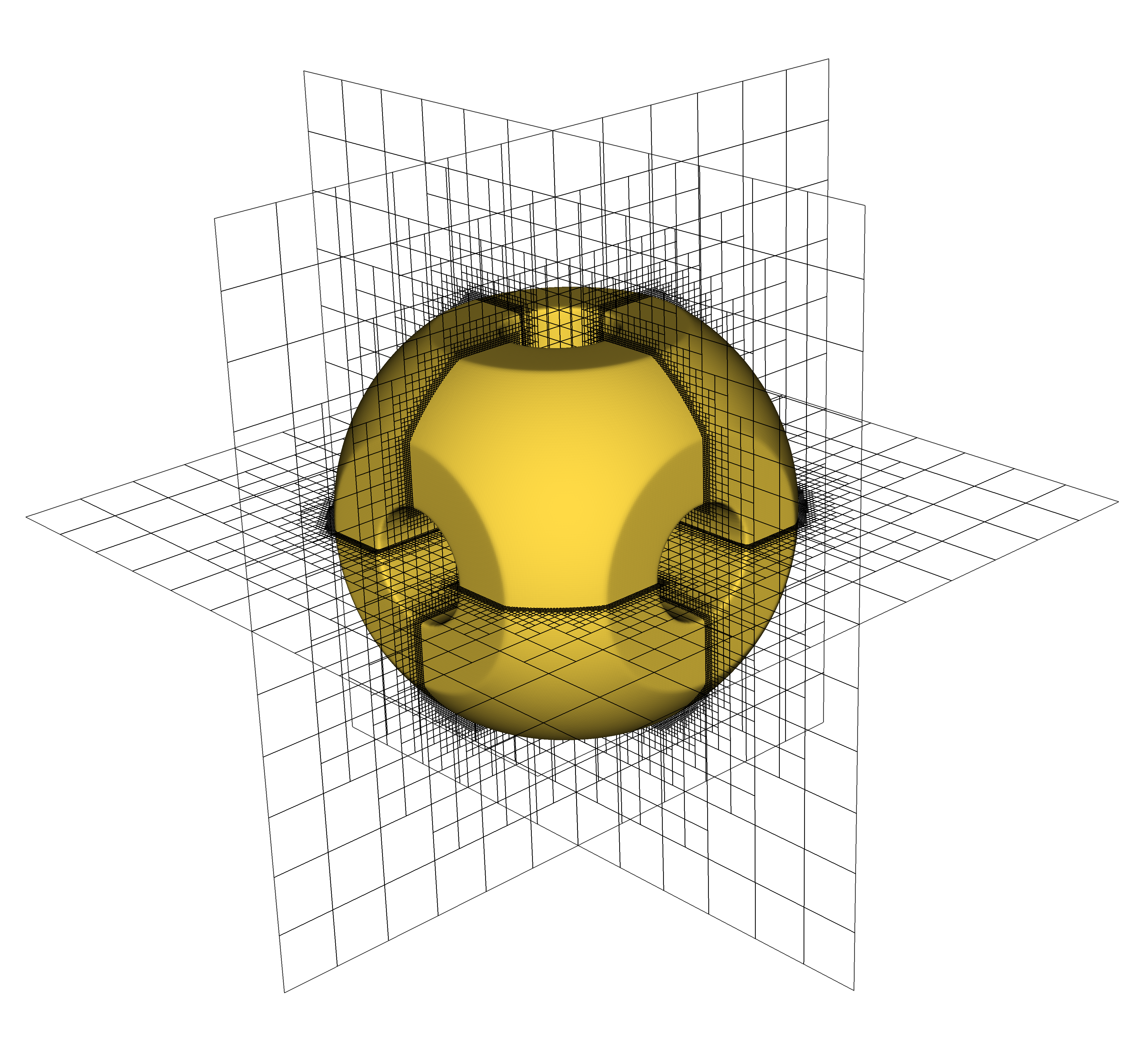} &
\includegraphics[height=0.38\textwidth]{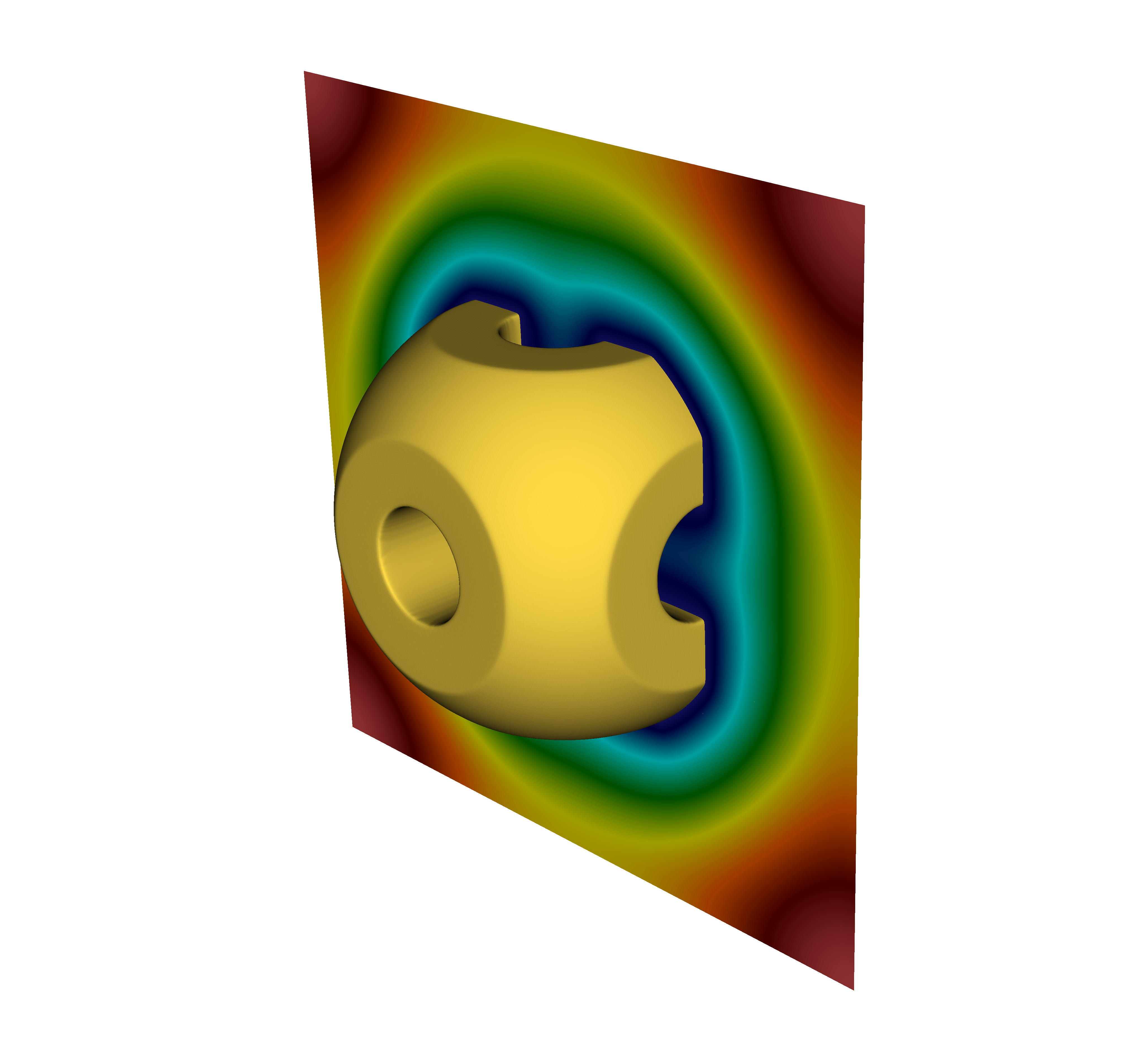} \vspace{-2mm}\\
\textrm{(a)} & \textrm{(b)} \\
\end{array}$
\end{center}
\vspace{-7mm}
\caption{(a) Adaptively refined background mesh used to model the target domain, and (b)
level set function computed using the distance from the zero level set of the
geometric primitive-based geometry.}
\label{fig_3D_geom_prim_bg}
\end{figure}

Using the approach outlined in Section \ref{sec_level_set_rep}, we
combine the geometric primitives on a third-order
source mesh that is adaptively refined five times around the zero level set of the target domain. We then compute the distance function on this background mesh using the $p-$Laplacian solver \cite{belyaev2015variational}. This distance function is used as the level set function in \eqref{eq_F_full_sigma}. Figure \ref{fig_3D_geom_prim_bg}(a) shows a slice-view of the background mesh along with the zero iso-surface of the level set function, and Fig. \ref{fig_3D_geom_prim_bg}(b) shows a slice-view of the level set function computed as the distance from the zero level set of the geometric primitive-based description of the domain. Our numerical experiments showed that using a third order source mesh with five adaptive refinements was computationally cheaper than using a lower order mesh with more adaptive refinements to obtain the same level of accuracy for capturing the the curvilinear boundary using the distance function.

\begin{figure}[tb!]
\begin{center}
$\begin{array}{cc}
\includegraphics[height=0.38\textwidth]{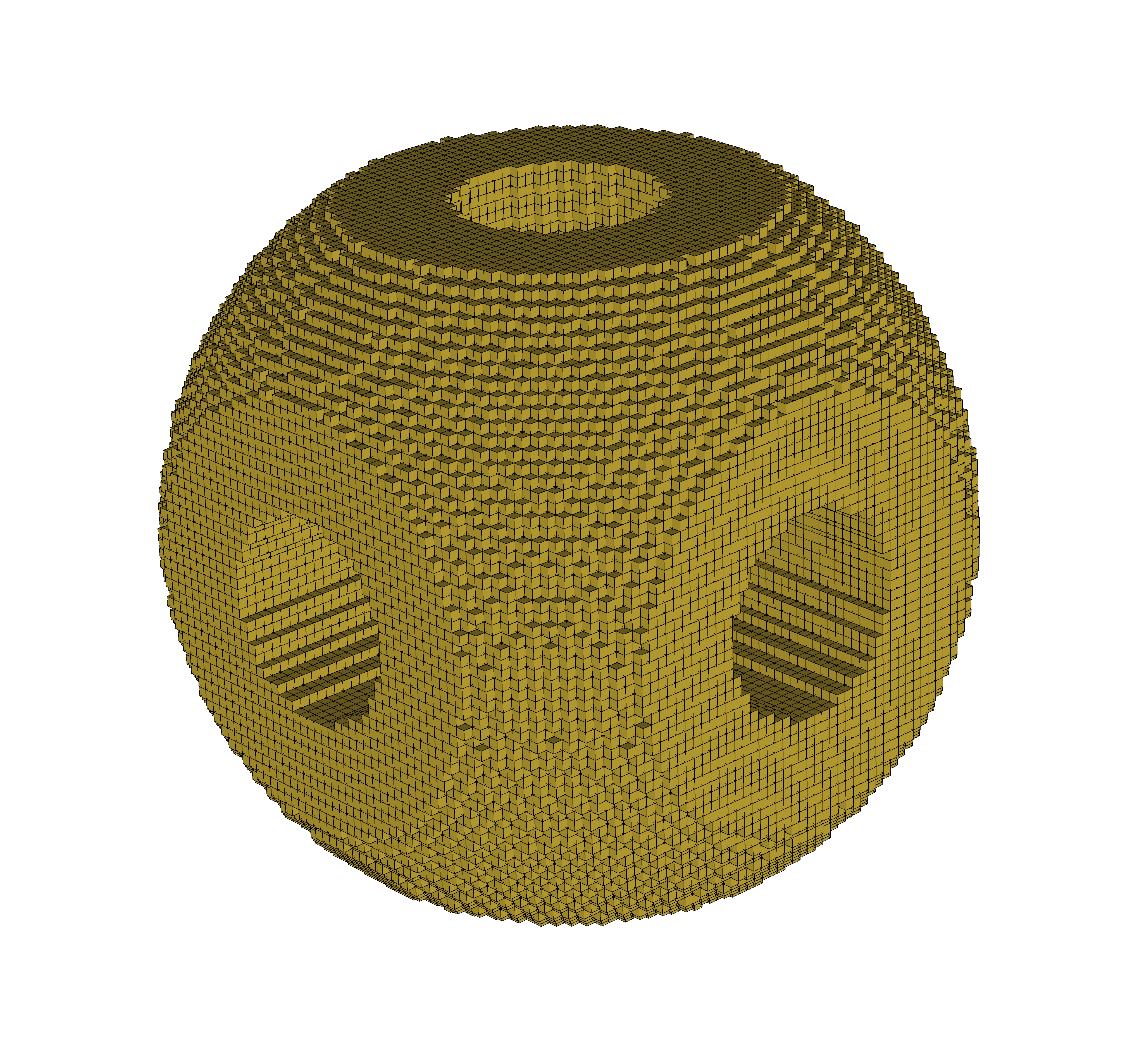} &
\includegraphics[height=0.38\textwidth]{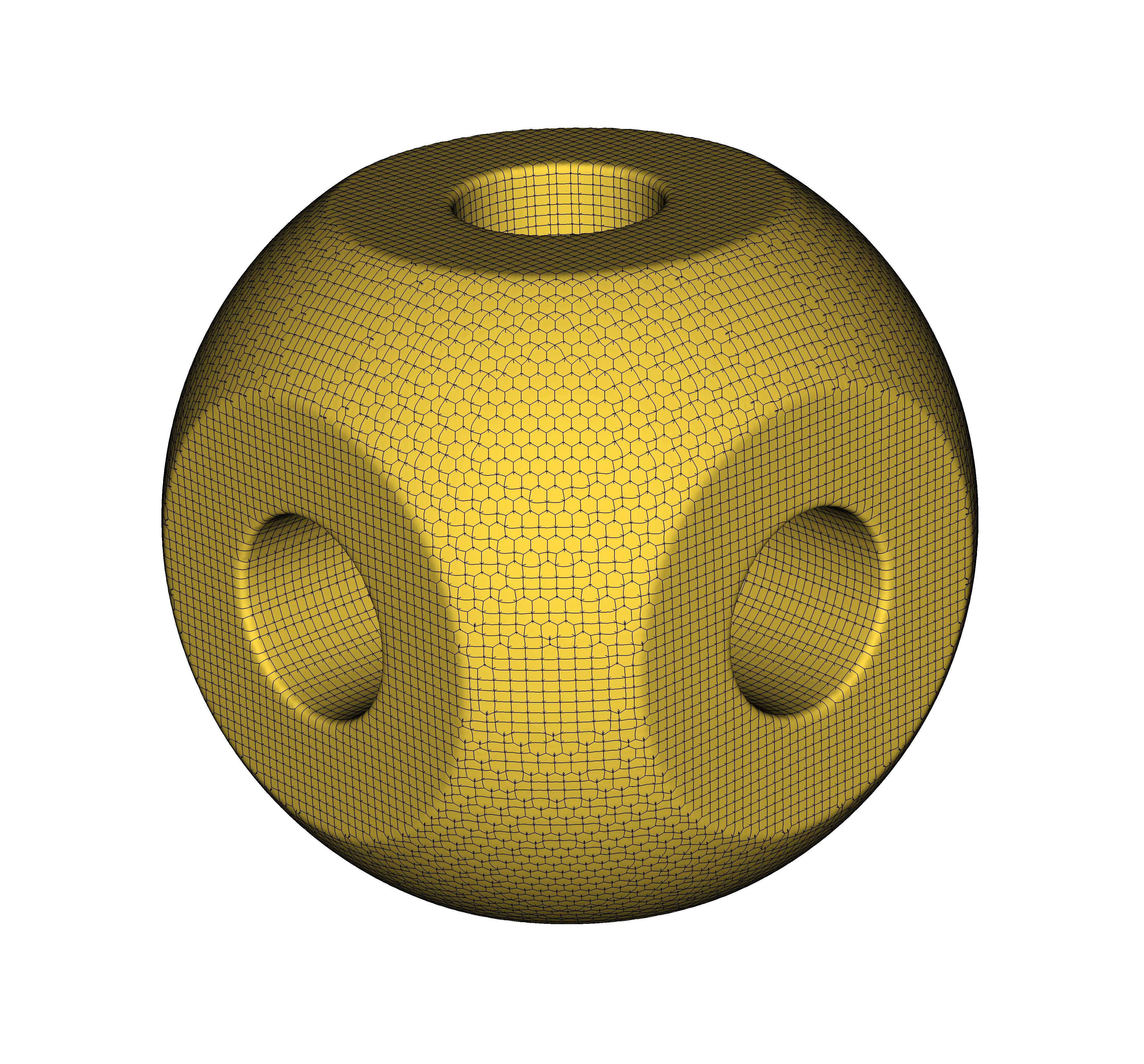}\vspace{-2mm} \\
\textrm{(a)} & \textrm{(b)} \\
\end{array}$
\end{center}
\vspace{-7mm}
\caption{(a) Initial trimmed mesh and (b) final fitted mesh for the primitive-based geometry test.}
\label{fig_3D_geom_prim_opt}
\end{figure}

The input mesh to be fit for this problem is a uniform second-order $128 \times 128 \times 128$ Cartesian-aligned mesh for $\Omega \in [0, 1]^3$.
\cblue{This mesh is trimmed by using \eqref{eq_eta} and removing all elements with $\eta_E = 0$. The minimum Jacobian in the trimmed mesh is $4.8\times 10^{-7}$}. \cgreen{The trimmed mesh is optimized using \eqref{eq_F_full_sigma} where the nodes on the boundary are set for alignment. Figure \ref{fig_3D_geom_prim_opt} shows the input and the optimized mesh, where the achieved fitting error is
$|\sigma|_{\mathcal{S},\infty} = \mathcal{O}(10^{-6})$ with the minimum Jacobian at the boundary decreasing to $4.8\times 10^{-10}$.}

\begin{figure}[b!]
\begin{center}
$\begin{array}{cc}
\includegraphics[height=0.40\textwidth]{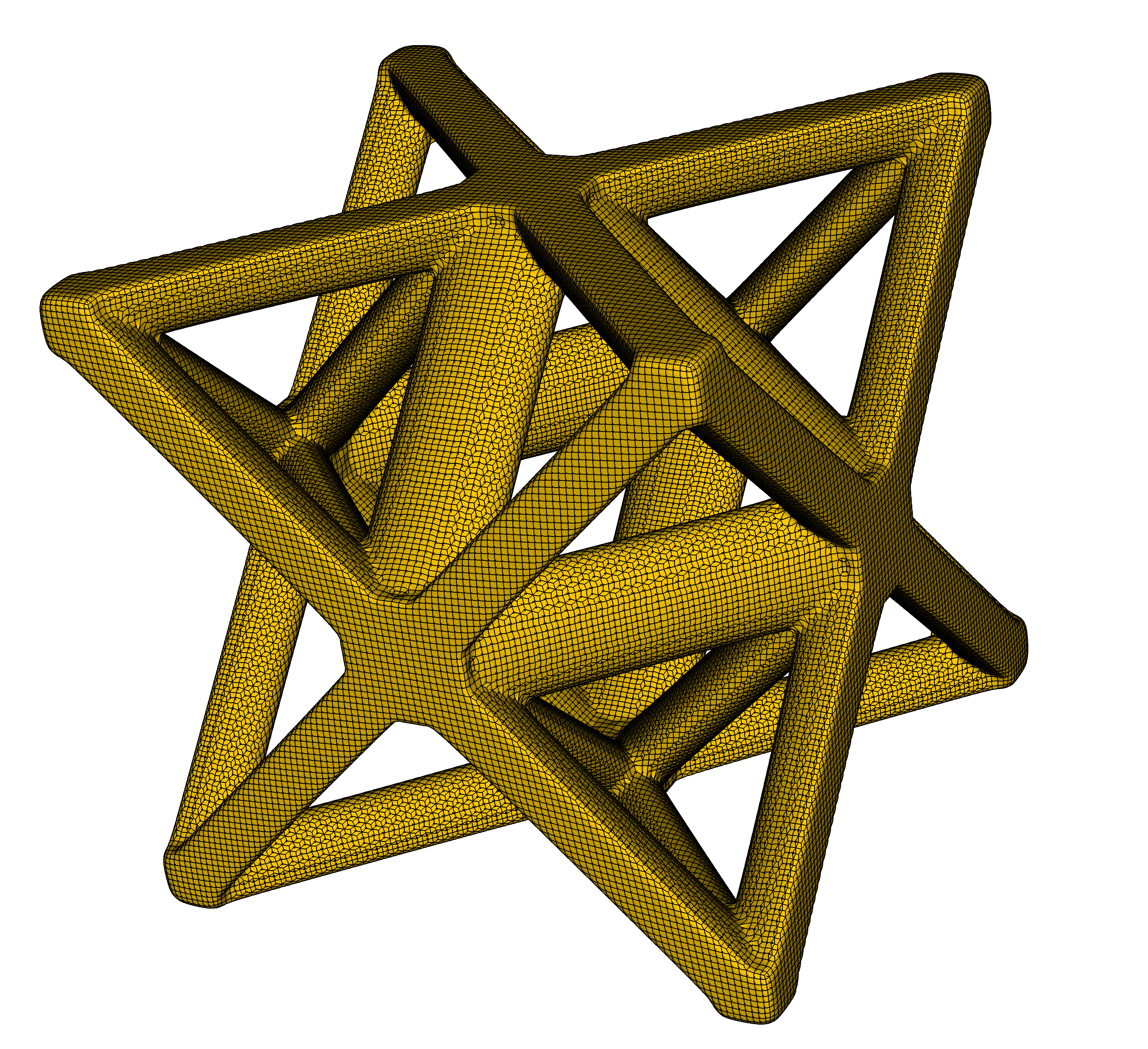} &
\includegraphics[height=0.40\textwidth]{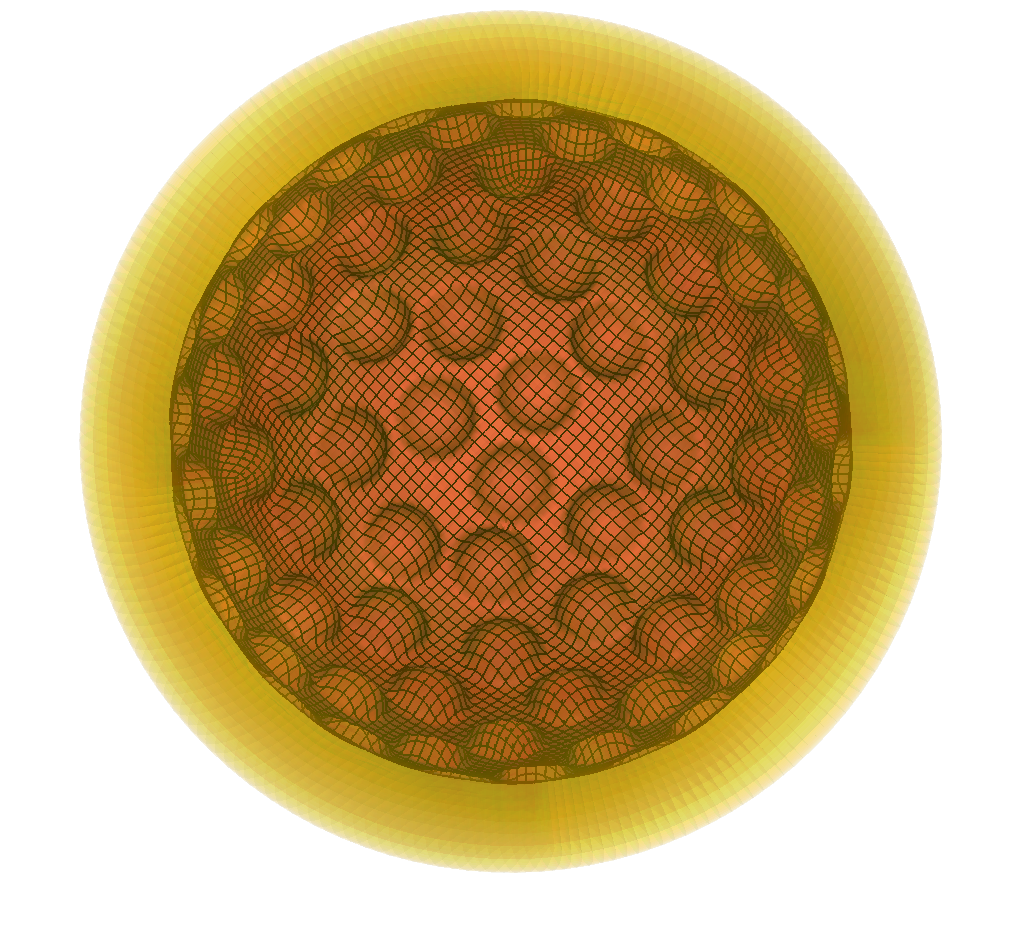}\vspace{-2mm} \\
\textrm{(a)} & \textrm{(b)} \\
\end{array}$
\end{center}
\vspace{-7mm}
\caption{\cred{(a) Boundary-fitted mesh generated using a geometric-primitive based description for an Octet truss. (b) Interface-fitted mesh for a multimaterial domain with concentric spherical shells with uniformly spaced indents of same size at the material interface. The inner shell is colored red and the outer shell is translucent and colored golden.}}
\label{fig_3D_apps}
\end{figure}

\cred{Other 3D applications of interest that are currently leveraging the proposed method are shown next. Figure \ref{fig_3D_apps}(a) shows an example of one of the boundary-fitted meshes obtained for
simulation and design of lattice structures in the context of additive manufacturing. Figure \ref{fig_3D_apps}(b) shows an example of one of the interface-fitted meshes obtained for analyzing the impact of fluid flow on parameterized surfaces. Here, the target multimaterial domain consists of two concentric shells with parameterized locations and sizes for indents at the material interface. Note, the example shown in Fig. \ref{fig_3D_apps}(b) consists of uniformly spaced indents of the same size. The inner shell is colored red and the outer shell is translucent and colored golden, with the mesh morphed to align to the indents highlighted at the interface.}

\subsection{Interface Fitting for Shape Optimization Application}
\begin{figure}[b!]
\begin{center}
$\begin{array}{c}
\includegraphics[width=0.6\textwidth]{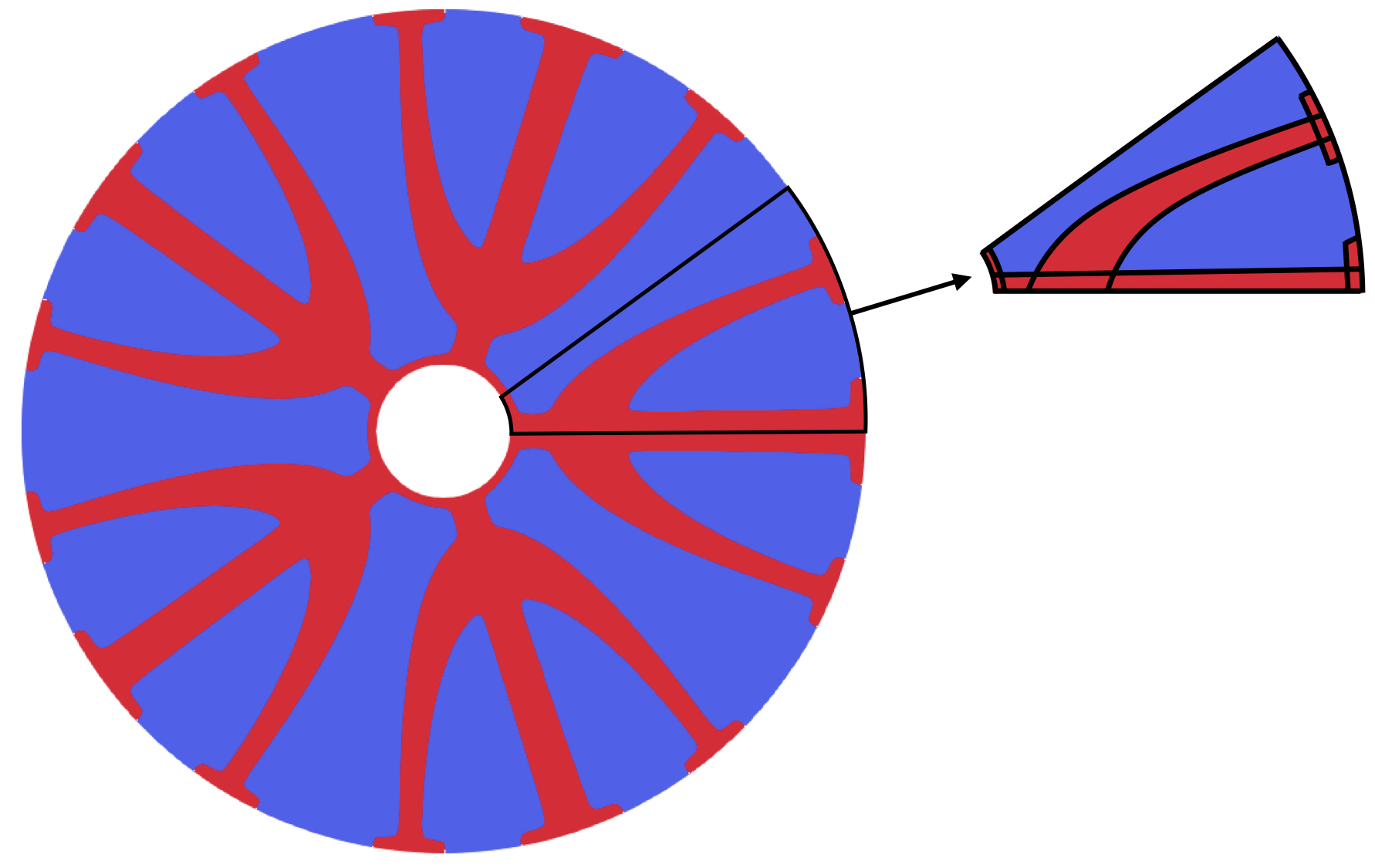}
\end{array}$
\end{center}
\vspace{-7mm}
\caption{Reactor domain to be meshed for shape optimization, along with the
symmetric portion and its primitive decomposition.
The internal material interfaces must be fitted in the final mesh.}
\label{fig_reactor_domain}
\end{figure}

This example serves to demonstrate the applicability of the proposed approach to setup the initial multimaterial domain to be used in a shape optimization problem.
Figure \ref{fig_reactor_domain} shows the cross section of a tubular reactor that consists of a highly conductive metal (\emph{red}) and a low conductivity heat generating region (\emph{blue}). \cblue{The design optimization problem is formulated such that the energy production in the system is maximized while keeping the overall volume of the red region constant}.
To achieve this, the shape of the \emph{red} subdomain is morphed using a gradient-based approach in our in-house design optimization framework that requires an interface fitted mesh as an input.
This initial fitted mesh is generated using the method described in this manuscript.

Exploiting the cyclic symmetry, we discretize a portion of the domain via a triangular mesh. Since this initial mesh does not need to align with the material interface, it can be conveniently generated using an automatic mesh generator \cite{blacker2016cubit}. We then use an approach similar to the previous section for interface fitting where the multimaterial domain is realized as a combination of geometric primitives (annulus, parabola, and trapezium,
as highlighted in Figure \ref{fig_reactor_domain}).
The finite element distance function from the interface is computed
\cgreen{using the p-Laplacian solver of \cite{belyaev2015variational}, Section 7,}
and used as the level-set function $\sigma(x)$ in \eqref{eq_F_full_sigma}.

\begin{figure}[tb!]
\begin{center}
$\begin{array}{cc}
\includegraphics[width=0.3\textwidth]{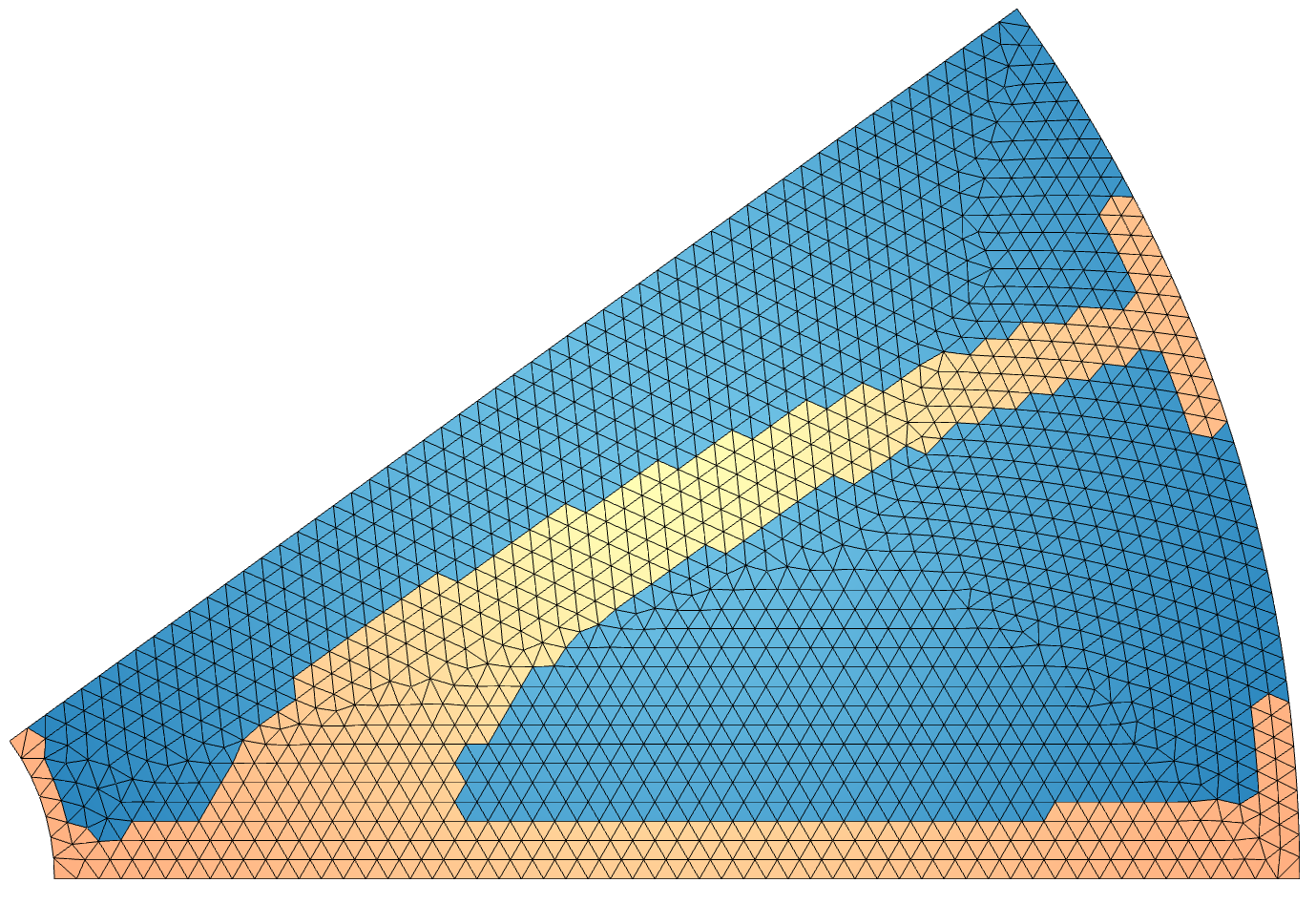} &
\includegraphics[width=0.3\textwidth]{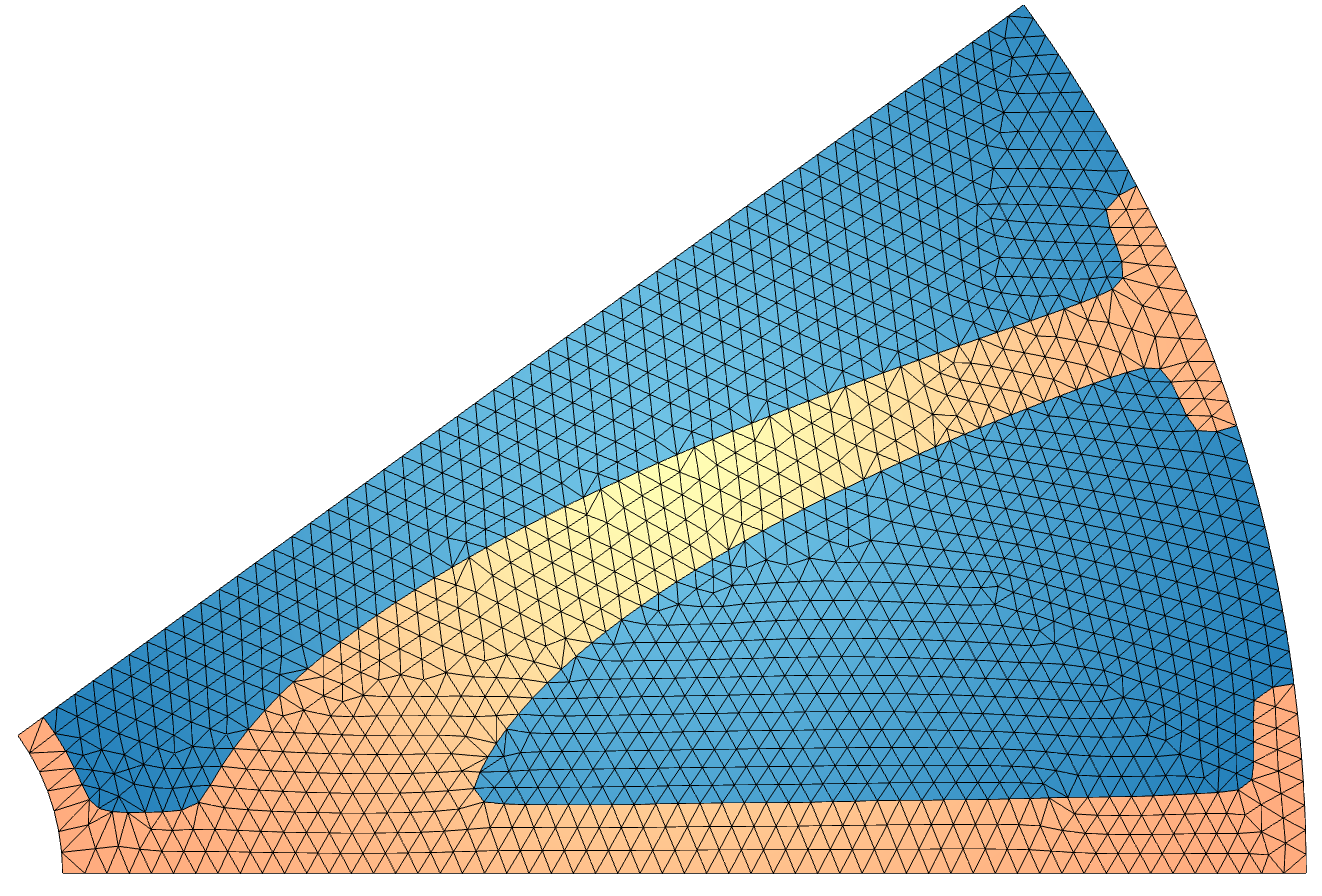} \vspace{-2mm} \\
\textrm{(a)} & \textrm{(b)} \\
\end{array}$
\end{center}
\vspace{-7mm}
\caption{(a) Original mesh and (b) interface fitted mesh
         for the reactor design problem.}
\label{fig_reactor_meshes}
\end{figure}

\begin{figure}[tb!]
\begin{center}
$\begin{array}{ccc}
\includegraphics[width=0.27\textwidth]{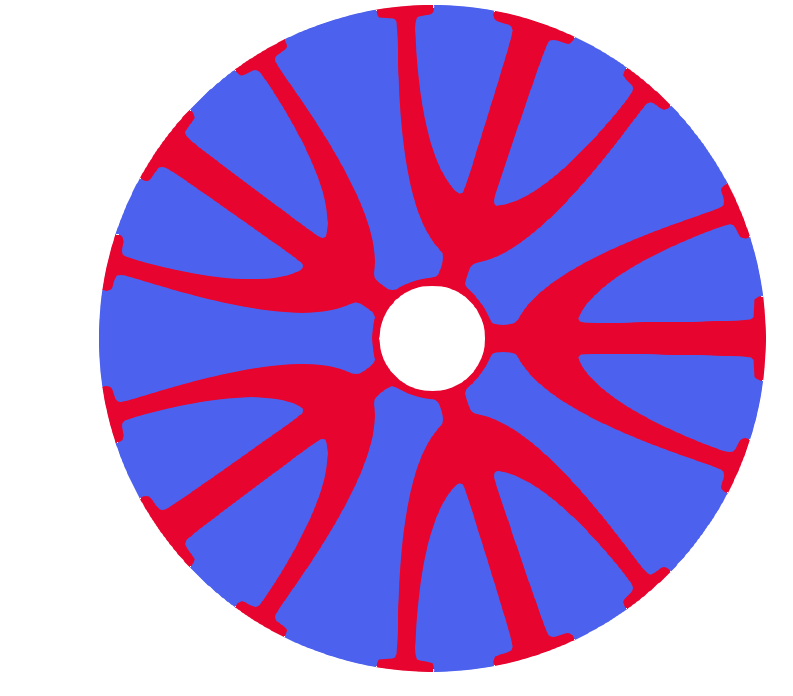} &
\includegraphics[width=0.27\textwidth]{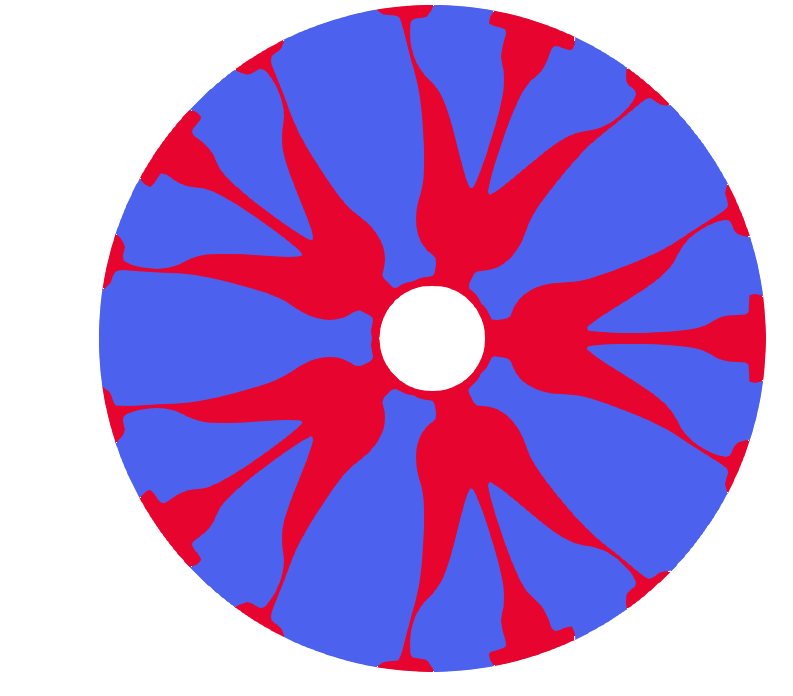} &
 \\
\text{(a)} &
\text{(b)} \\
\includegraphics[width=0.27\textwidth]{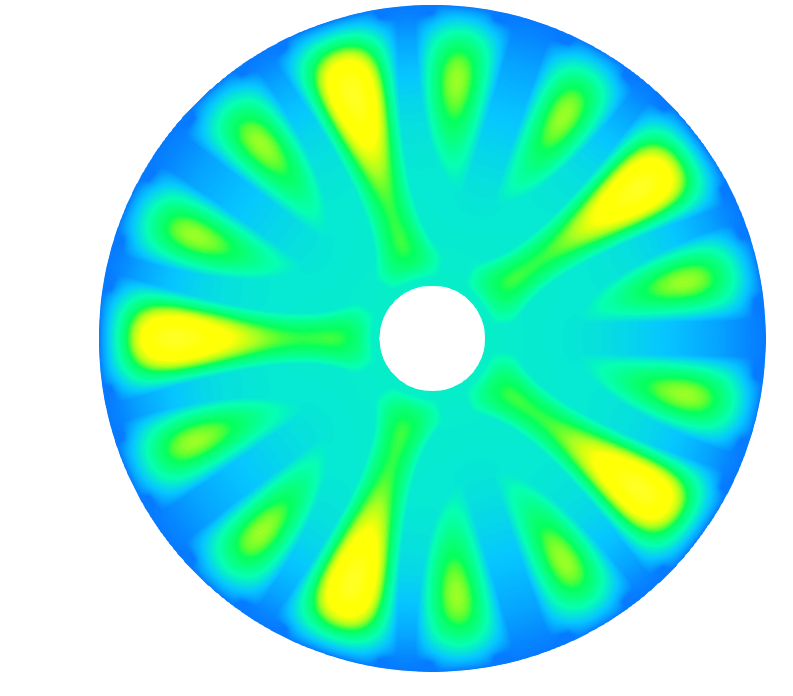} &
\includegraphics[width=0.27\textwidth]{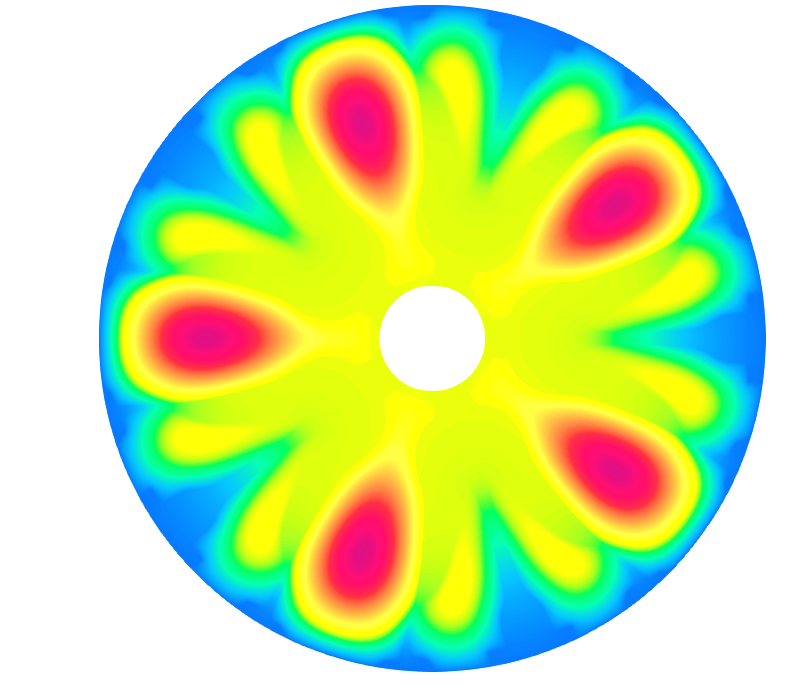} & \hspace{-10mm}
\includegraphics[width=0.1\textwidth,height=0.2\textwidth]{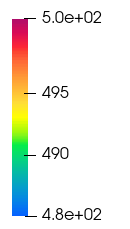} \\
\text{(c)} &
\text{(d)}
\end{array}$
\end{center}
\vspace{-7mm}
\caption{(a) Initial fitted domain obtained using the proposed method and (b) shape optimized domain from our in-house design optimization framework. Corresponding \cblue{temperature (Kelvin)} for the (c) initial domain and (d) shape optimized domain.}
\label{fig_reactor_heat}
\end{figure}

Figure \ref{fig_reactor_meshes}(a) shows the input non-fitted mesh, colored by
the adaptively assigned material indicator, such that there is at most one
face per triangle marked for fitting.
Figure \ref{fig_reactor_meshes}(b) shows the adapted mesh which
aligns with the material interface of the domain. The maximum interface fitting
error of the optimized mesh is $\mathcal{O}(10^{-10})$, \cgreen{and the minimum Jacobian in the mesh decreases from $10^{-7}$ to $2\times10^{-8}$ at the nodes along the interface due to mesh deformation enforcing alignment to the target surface.}
Finally, Figures \ref{fig_reactor_heat}(a)-(d) show the initial
interface-fitted domain and the shape optimized domain along with
the corresponding \cblue{temperature (Kelvin)} fields. As we can see, the proposed method is effective for generating high quality interface fitted meshes with minimal user intervention, and is currently being used for similar 3D shape optimization applications that will be presented in future work.

\section{Conclusion \& Future Work}
\label{sec_concl}
We have presented a novel method to morph and align high-order meshes to the
domain of interest.
We formulate the mesh optimization problem as a variational minimization of the sum of a chosen mesh-quality metric and a penalty term that weakly forces the selected faces of the mesh to align with the target surface. The penalty-based formulation makes the proposed method suitable for adoption in existing mesh optimization frameworks.

\cblue{There are three key features of the proposed method that enable its robustness.
First, a source mesh is used to represent the level set function with sufficient accuracy when the mesh being morphed does not have enough resolution or is beyond the target surface (Section \ref{sec_level_set_rep}). Second, an adaptive approach is proposed for setting the fictitious material indicators in the mesh to ensure that the resulting material interface can align to the target surface for interface fitting (Section \ref{sec_marking}). Finally, an adaptive approach for setting the penalization weight is developed to eliminate the need for tuning the penalization weight on a case-by-case basis (Section \ref{sec_adaptive_sigma}). Numerical experiments demonstrate that the proposed method is effective for generating boundary- and interface-fitted meshes for nontrivial curvilinear geometries.}

In future work, we will improve the method by developing mesh refinement strategies for hexahedral and tetrahedral meshes, which are required when the mesh topology limits the fit of a mesh to the target surface (Section \ref{sec_marking}).
We will also explore ways for aligning meshes to domains
with sharp features \cite{ohtake2002dual,zahr2020implicit}, as we
currently assume that the level set function $\sigma$ used in
\eqref{eq_F_full_sigma} is sufficiently smooth around its zero level set.
Finally, we will also look to optimize our method and leverage
accelerator-based architectures by utilizing partial assembly and
matrix-free finite element calculations \cite{camier2022accelerating}.

\bibliographystyle{elsarticle-num}
\bibliography{bif}


\end{document}